\documentclass[fleqn,usenatbib]{mnras}

\usepackage{newtxtext,newtxmath}

\usepackage[T1]{fontenc}
\usepackage{ae,aecompl}

\usepackage{graphicx}	
\usepackage{amsmath}	
\usepackage{amssymb}	

\title[BCG colours]{The Evolution of Brightest Cluster Galaxies in the Nearby Universe I: Colours and Stellar Masses from the Sloan Digital Sky Survey and Wide Infrared Survey Explorer}

\author[P. Cerulo et al.]{
P. Cerulo,$^{1}$\thanks{E-mail: pcerulo@astro-udec.cl}
G. A. Orellana,$^{2}$
and G. Covone$^{3,4}$
\\
$^{1}$Department of Astronomy, Universidad de Concepci\'{o}n, Casilla 160-C, Concepci\'{o}n, Chile\\
$^{2}$Instituto de F\'isica y Astronom\'ia, Universida de Valpara\'iso, Avda. Gran Breta\~na 1111, Valpara\'iso, Chile.\\
$^{3}$Dipartimento di Fisca ``E. Pancini'', University of Naples ``Federico II'', Naples, Italy \\
$^{4}$INFN, Sezione di Napoli,  Naples, Italy
}

\date{Accepted XXX. Received YYY; in original form ZZZ}

\pubyear{2015}

\hypersetup{draft}
\begin{document}
\label{firstpage}
\pagerange{\pageref{firstpage}--\pageref{lastpage}}
\maketitle

\begin{abstract}
We present a study of the evolution of brightest cluster galaxies (BCGs) in a sample of clusters at $0.05 \leq z<0.35$ from the SDSS and WISE with halo masses in the range $6 \times 10^{13}M_\odot$ (massive groups) - $10^{15.5}M_\odot$ (Coma-like clusters). 
We analyse optical and infrared colours and stellar masses of BCGs as a function of the mass of their host haloes. We find that BCGs are mostly red and quiescent galaxies and that a minority ($\sim 9$\%) of them are star-forming. We find that the optical $g-r$ colours are consistent with those of red sequence galaxies at the same redshifts; however, we detect the presence of a tail of blue and mostly star-forming BCGs preferentially located in low-mass clusters and groups. Although the blue tail is dominated by star-forming galaxies, we find that star-forming BCGs may also have red $g-r$ colours, indicating dust-enshrouded star formation. The fraction of star-forming BCGs increases with redshift and decreases with cluster mass and BCG stellar mass. We find that cool-core clusters host both star-forming and quiescent BCGs; however, non cool-core clusters are dominated by quiescent BCGs. Star formation appears thus as the result of processes that depend on stellar mass, cluster mass and cooling state of the intra-cluster medium. Our results suggest no significant stellar mass growth at $z<0.35$, supporting the notion that BCGs had accreted most of their mass by $z = 0.35$. Overall we find a low (1\%) AGN fraction detected at IR wavelengths.
\end{abstract}




\section{Introduction}

Brightest cluster galaxies (BCG) are the most massive galaxies in our Universe and constitute a peculiar family of objects. They are located near the centres of their host clusters (although see \citealt{Lidman_2013} for cases of BCGs located away from the cluster centroid), have extended light profiles (cD, see \citealt{Tonry_1987}, \citealt{Graham_1996}) and little or no ongoing star formation (\citealt{Donahue_2010}, \citealt{Oliva_Altamirano_2014}, \citealt{Fraser_McKelvie_2014}, \citealt{Gozaliasl_2016}). Their stellar masses correlate with the halo mass of the host cluster or group (\citealt{Brough_2008}, \citealt{Lidman_2012}, \citealt{Oliva_Altamirano_2014}, \citealt{Bellstedt_2016}, \citealt{Lavoie_2016}, \citealt{Gozaliasl_2016}, although see \citealt{Zhao_2015b} for a different interpretation of this relationship). Although most observational studies agree in finding a factor of $\sim 2$ growth in stellar mass with redshift since at least $z \sim 1$ (\citealt{Aragon_Salamanca_1998}, \citealt{Lidman_2012}, \citealt{Lin_2013}, \citealt{Ascaso_2014}, \citealt{Shankar_2015}, \citealt{Bellstedt_2016}, \citealt{Lavoie_2016}), mainly through dry mergers (\citealt{Lidman_2013}, \citealt{Lavoie_2016}) with, in particular, a fast growth in stellar mass at redshifts $z > 0.5$ and a slower growth, less than a factor 0.5, at $z<0.5$ (\citealt{Lin_2013}, \citealt{Oliva_Altamirano_2014}), other authors find negligible stellar mass growth in the last 8 billion years (\citealt{Stott_2008}, \citealt{Whiley_2008}; see \citealt{Lidman_2012} for a discussion on the origin of the lack of detection of stellar mass growth). Their luminosity function deviates from a Schechter function (\citealt{Schechter_1976}, \citealt{De_Filippis_2011}, \citealt{Wen_2015b}) and is better fitted by a Gaussian, while their sizes grow by a factor of 2.5 since $z \sim 1$ (\citealt{Shankar_2015}).

Theoretical models predict that BCGs assemble most of their stellar mass through dry mergers (\citealt{Dubinski_1998}, \citealt{De_Lucia_and_Blaizot_2007}), with major mergers dominating in the early phases of their formation and minor mergers dominating at later stages of their evolution (\citealt{De_Lucia_and_Blaizot_2007}), when these galaxies grow their mass accreting smaller objects (galactic cannibalism) that fall into the cluster and lose velocity through dynamical friction ({\citealt{White_1976}}). \citealt{Shankar_2015} show that only models with little stellar stripping and low orbital energies of satellite galaxies can be able to reproduce the observed stellar mass and size growths. 

Simulations and semi-analytic models based on the hierarchical growth of structures (e.g.\ \citealt{Bower_2006}, \citealt{Croton_2006}, \citealt{Gabor_2012}) predict that BCGs should have optical colours that are on the blue side of the of the red sequence. This is not found in all observational works. 
For instance, in \cite{Valentinuzzi_2011} it can be seen that BCGs in the WIde-field Nearby Galaxy Cluster Survey (WINGS, \citealt{Fasano_2006}) have optical colours that are bluer than those that would be expected from the linear fit to the red sequence, but \cite{Bernardi_2007} show that the colours of BCGs are those typical of red sequence galaxies. 

The models of \cite{Tonini_2012} show that BCGs should have star formation down to low redshifts. Although most works show that BCGs in the local universe are mostly passive (e.g.:\ \citealt{Fraser_McKelvie_2014}), at high redshifts \cite{Webb_2015} and \cite{Bonaventura_2017} show an increase in the star formation activity of BCGs. \cite{Webb_2017} also report a case of a BCG with a significant amount of molecular gas in a cluster at $z=1.8$. Although star formation is rare in low-redshift BCGs, the works of \cite{Donahue_2010} and \cite{Liu_2012} suggest that recent or ongoing star formation in BCGs is related to the presence of cooling flows in clusters (see also \citealt{Hu_1985}, \citealt{Bildfell_2008}, \citealt{Cavagnolo_2008}, \citealt{Pipino_2009}, \citealt{Donahue_2015}, \citealt{Loubser_2016}).

\cite{Best_2007} report that radio-loud active galactic nuclei (AGN) are more abundant in BCGs than in other galaxies in clusters at $0.02 < z < 0.16$ in the Sloan Digital Sky Survey (SDSS, \citealt{York_2000}), while \cite{Croft_2007} show that the fraction of radio-loud AGN increases with the stellar mass of BCGs at redshifts $0.03 < z < 0.3$ in the SDSS, in agreement with the findings of \cite{Yuan_2016}, who studied BCGs up to $z \sim 0.45$ in the SDSS. Interestingly. \cite{von_der_Linden_2007} show that while BCGs are more likely than other galaxies to host radio-loud AGNs, they are less likely to show an optically detected AGN at their centre.

BCGs in the local Universe host very old stellar populations, consistent with formation redshifts $z > 2$ (\citealt{Brough_2007}, \citealt{von_der_Linden_2007}, \citealt{Whiley_2008}, \citealt{Loubser_2009}). However, while \citealt{Brough_2007} also find that the age of BCGs and Brightest group galaxies (BGG) is correlated with the X-ray luminosity of galaxy clusters and groups, a proxy for their halo mass, \citealt{Loubser_2009} do not find such a correlation. The samples of \cite{Loubser_2009} and \cite{Brough_2007} were small, with only 49 objects in the first work an 6 objects in the second, which does not make possible to draw statistically significant conclusions on the existence of a correlation between stellar age and halo mass.

In the present paper we study BCGs in a large sample of clusters drawn from the SDSS. We address the problem of the relationships between BCG properties and the properties of their hosting clusters to study the links between their formation and evolution and the assembly of the cosmic structures that contain them. This is the first in a series of papers devoted to the study of the physics of brightest cluster galaxies in the nearby Universe. In this work we present the sample and focus on the evolution of colour, stellar mass and star-formation activity in nearby BCGs. Detailed andlyses of star formation and AGN activity will be presented in papers that are currently in preparation.

The paper is organized as follows: we present the data in Section 2 and the analysis and results in Section 3. Section 4 discusses the results and in Section 5 we summarise our main conclusions. Throughout the paper we use a cosmology with $\Omega_m = 0.27$, $\Omega_\Lambda = 0.73$, and $H_0 = 70.5$ km $\cdot$ s$^{-1}$ $\cdot$ Mpc$^{-1}$ (\citealt{Hinshaw_2009}), unless otherwise stated. SDSS magnitudes are quoted in the AB system, while WISE magnitudes and fluxes are referred in the Vega system unless otherwise stated. We will interchangeably use the symbols $z$ and $z_{phot}$ to refer to photometric redshifts and $z_{spec}$ to refer to spectroscopic redshifts. We will use the notation $R_{200}$ to indicate the radius within which the local density is 200 times the critical density of the Universe at the redshift of each BCG and $M_{200}$ to indicate the total mass enclosed within this radius.

\section{Data}

\subsection{The Cluster Catalogue} \label{cluster_catalogue}

We use the catalogue of galaxy clusters presented in \cite{Wen_2012} in the updated version of \cite{Wen_2015a} (hereafter WHL15). The initial version of the catalogue was composed of 132,684 clusters detected in the SDSS Data Release 8 (DR8, \citealt{Aihara_2011}) at redshifts $0.05 < z < 0.8$. The catalogue was updated by \cite{Wen_2015a} using the better spectroscopic information available in the SDSS Data Release 12 (DR12, \citealt{Alam_2015}). The position of each cluster in the catalogue corresponds to the position of its BCG. The catalogue is 80\% complete at $z<0.4$ for clusters with halo masses $M_{200} > 6.0 \times 10^{13}M_\odot$, with the completeness dropping sharply beyond this redshift (see Figure 6 in \citealt{Wen_2012}).
Details on the estimation of the cluster halo masses are given in Section 3. 

We summarize here the cluster detection algorithm which is explained in detail in \cite{Wen_2012}. For each galaxy the number $N_{0.5}$ of neighbours with $M_r < -21.5$ mag, where $M_r$ is the $r$-band absolute magnitude, in a cylinder with radius 0.5 Mpc and length $0.08\times(1+z)$, centred in the spatial and redshift position of the galaxy, is estimated. A friend-of-friend algorithm (\citealt{Huchra_Geller_1982}) is run to search the galaxies linked to the central one within a spatial distance of 0.5 Mpc and a redshfit distance of $0.08\times(1+z)$. The linked galaxy with the highest $N_{0.5}$ is defined as the temporary centre of a candidate cluster. In the case in which two or more galaxies have the same maximum number, the brightest one is taken as the central galaxy of the candidate cluster. For each candidate cluster the number of galaxies with $M^e_{r}(z) < -20.5$ mag, where $M^e_{r}(z)$ is the evolution corrected absolute magnitude in the $r$ band, within a projected distance of $1.0$ Mpc and a photometric redshfit difference $\Delta z_p = \pm 0.04\times(1+z)$ is taken as the number of member galaxies. 
The median of the photometric redshifts of these galaxies is taken as the photometric redshift of the candidate cluster.

The BCG is defined as the brightest of the members of each candidate cluster, and the centre of the cluster is reset to the position of the BCG. 
Only candidate clusters with richness $R_{L^*} > 12$ are retained in the catalogue. 
\cite{Wen_2012} define the richness  as the ratio between the total luminosity of member galaxies within $R_{200}$ and the turn-off luminosity of the \cite{Schechter_1976} function $L^*$.

\cite{Wen_2012} checked their candidate cluster catalogue against a previous version of their catalogue (\citealt{Wen_2009}) and other cluster catalogues built from SDSS (e.g.\ AMF, GMBCG, maxBCG), as well as with clusters from the Abell catalogue that fall in the SDSS area. They found that the agreement with those catalogues varied between 40 and 90\% at $z < 0.4$, being higher at lower redshifts. 
On the high-mass end,  \cite{Sereno2017} found that $\sim $ 75\% of Sunyaev-Zel'dovich-selected clusters have a counterpart in the WHL15 catalogue.
\cite{Wen_2012} also visually inspected their candidate clusters to search for false detections and removed about 5,000 systems from the catalogue.

The WHL15 catalogue reports the positions and photometric redshifts of all the clusters which, by definition, correspond to those of their BCGs. We selected only clusters at $z < 0.35$ with halo masses $M_{200} > 6.0 \times 10^{13}M_\odot$. The sample obtained in this way contains 74,275 BCGs.

\subsection{Optical Data: the SDSS DR12} \label{optical_data}

We retrieved the photometric data in the optical $u$, $g$, $r$, $i$, and $z$ bands from the SDSS DR12 database. Objects were queried within a radius of $\sim 0.02''$ from the position of each BCG in the WHL15 catalogue using the scalar function {\ttfamily{fGetNearestObjEq}}.

We used SDSS model magnitudes ({\ttfamily{modelMag}}), corrected for Galactic extinction according to the maps of \cite{Schlegel_1998}, to estimate colours, while we used the Petrosian magnitude in the $r$ band as the definition of total magnitude in our analysis. Model magnitudes are estimated by first fitting a \cite{DeVauc_1948} and an exponential profiles to each galaxy in the $r$ band and then choosing the model with the highest likelihood between the two. The highest-likelihood model is then applied to the other filters after convolving with their respective PSFs, and magnitudes are estimated within the half-light radius of the model. As explained in the SDSS support documentation\footnote{\url{https://www.sdss.org/dr12/algorithms/magnitudes/}}, model magnitudes provide the most unbiased estimates of galaxy colours.

We also downloaded photometric redshifts (photo-z, $z_{\rm phot}$) and rest-frame photometry from the {\ttfamily{Photoz}} table in the SDSS database. Details on the derivation of photometric redshifts for the SDSS DR12 are in \cite{Beck_2016}, and we refer the reader to that paper. Photo-z where obtained with a hybrid technique consisting of two subsequent steps: a local linear regression that used the SDSS spectroscopic sample as a training set for the estimation of the photometric redshifts, and a spectral energy distribution (SED) fitting algorithm that used the photo-z estimated in the first step to find the best fitting template to the SDSS optical $ugriz$ photometry. With this technique \cite{Beck_2016} can achieve an accuracy, parametrised with the standard deviation of the normalised redshift estimation error $z_{\rm norm} = (z_{\rm phot}-z_{\rm spec})/(1+z_{\rm spec})$, where $z_{\rm spec}$ is the spectroscopic redshift, of 0.0205.

We selected BCGs with redshifts in the interval $0.05 \leq z_{\rm phot} < 0.35$ with $z_{\rm phot}$ errors less than 0.03, number of nearest neighbours in the local linear regression fit greater than 95, $\chi^2 < 6.0$, and photometric error class 1. We also rejected objects that have {\ttfamily{modelMag}} fractional errors greater than 10\%. The sample after this selection is composed of 18,706 clusters.

\subsection{Infrared Data: WISE} \label{IR_data}

We complemented our optical data with near and mid infrared (IR) data from the Wide Infrared Survey Explorer (WISE, \citealt{Wright_2010}). WISE is an IR satellite that mapped the entire sky in four IR filters, namely $W1$ ($\rm 3.6 \mu m$), $W2$ ($\rm 4.3 \mu m$), $W3$ ($\rm 12 \mu m$), and $W4$ ($\rm 22 \mu m$)\footnote{Although \cite{Brown_2014} showed that the W4 filter is centred at $\lambda=22.8 \mu m$, we will use 22$\mu m$ as the definition of the central wavelength of the W4 filter in the remaining of the paper.}.

We selected galaxies in the AllWISE Source Catalog \footnote{http://irsa.ipac.caltech.edu/cgi-bin/Gator/nph-dd}, searching objects within a radius of 1$''$ from the WHL15 BCGs. Following the WISE Explanatory Supplement (see also \citealt{Fraser_McKelvie_2014}), we selected galaxies with profile-fitting magnitudes ({\ttfamily{promag}}), adopting the following criteria:
\begin{itemize}
\itemsep=-1.2em
\item (1) $S/N > 5$; \\
\item (2) Reduced $\chi^2$ of profile fitting larger than zero; \\
\item (3) $\rm w?\_flg \leq 2$ (the ? symbol represents the values 1-4), implying no upper limits, non-zero bit flag tripped, not all pixels unusable, and no saturation. This cut does not prevent, however, that another source may fall within the measurement aperture of the primary source and the presence of few bad pixels; \\ 
\item (4) Sources unaffected by known artefacts (parameter $\rm cc\_flags = 0$).
\end{itemize}

We found 114,720 BCGs detected in both W1 and W2, 6,077 BCGs that are simultaneously detected in $W1$, $W2$ and $W3$ and 632 BCGs that are detected in $W1$, $W2$, $W3$ and $W4$. 

After applying our selection in photometric redshift and the cuts in photometric redshift quality, photometry quality and halo mass described in Section 2.2, we end up with a cluster halo-mass complete sample of  BCGs with reliable optical photometry and $z_{\rm phot}$, of which 17,460 have photometry in $W1$, 17,857 have photometry in $W2$, 2,132 have photometry in $W3$, and 174 have photometry in $W4$. BCGs with photometry in $W1$ and $W2$ amount to a total of 16,716: for these galaxies we can determine their $W1-W2$ colours. There are finally 1,857 BCGs that we can classify as quiescent, star-forming or AGN hosts according to their $W1-W2$ and $W2-W3$ colours (see Section \ref{star_forming_BCG}). The magnitude distributions in all the WISE bands before and after the application of the cuts in optical photometry quality, photometric redshift quality and halo mass are shown in Appendix A for the BCGs at $0.05 \leq z_{phot} < 0.35$.

Because the {\ttfamily{promag}} algorithm is optimized for point sources, this measurement of the flux underestimates the total flux of extended sources. To mitigate this \citet{Chang_2015} proposed an empirical correction for the filters $W1$, $W2$ and $W3$, using the effective radii in the SDSS r-band:
\begin{equation}
\Delta m=0.10 + 0.46\log(\rm R_e) + 1.47\log(\rm R_e)^2 + 0.08\log(\rm R_e)^3,
\end{equation}
where $\rm R_e$ is the SDSS DR12 r-band effective radius in arcseconds corrected downward by a factor of 1.5$ \pm 0.2''$ to account for the differences between optical and IR galaxy effective radii. We used the {\ttfamily{fracDeV\_r}} parameter to decide whether to adopt half-light radii from \cite{DeVauc_1948} or exponential fits to the galaxies as effective radii of each BCG. More precisely, we adopted \cite{DeVauc_1948} half-light radii if {\ttfamily{fracDeV\_r}}$\geq 0.5$ and exponential half-light radii if {\ttfamily{fracDeV\_r}}$< 0.5$. We applied the correction to all BCGs with $\rm R_e \geq 0.5''$, while for galaxies with $\rm R_e < 0.5''$ the {\ttfamily{promag}} magnitudes remained unchanged. We did not apply any empirical correction to the $W4$ filter.

\section{Analysis and Results} \label{results}

\subsection{Cluster Sub-samples} \label{cluster_subsamples}

The aim of this paper is to relate the properties of BCGs to their host clusters. For this reason we divided the WHL15 sample into three subsets according to the mass of their hosting haloes. We used the scaling relation presented in Equation 2 of \cite{Wen_2012} between cluster richness and halo mass $M_{200}$ of the clusters. We use here as definition of halo mass the dark matter mass within $R_{200}$. The scaling relation of \cite{Wen_2012} was confirmed by \cite{Covone_2014} by means of a weak lensing analysis. We used the uncertainties on the parameters of Equation 2 of \cite{Wen_2012} to estimate the errors on $M_{200}$ through Monte-Carlo simulations.

To define the cluster sub-samples, we considered dark matter haloes with masses at $z=0$, $M_{200} = 10^{14}M_\odot$ and $M_{200} = 10^{14.7}M_\odot$ and evolved them back to redshift $z=0.35$ to reconstruct their accretion histories. We used the {\ttfamily{commah}} {\ttfamily{python}} library which derives the mass accretion history of haloes using the extended Press-Schechter formalism (\citealt{Correa_2015a,Correa_2015b,Correa_2015c})\footnote{{\ttfamily{commah}} is available at \url{https://github.com/astroduff/commah}.}.

\begin{figure}
\centering
\includegraphics[width=0.5\textwidth, trim=0.0cm 32.0cm 0.0cm 0.0cm, clip]{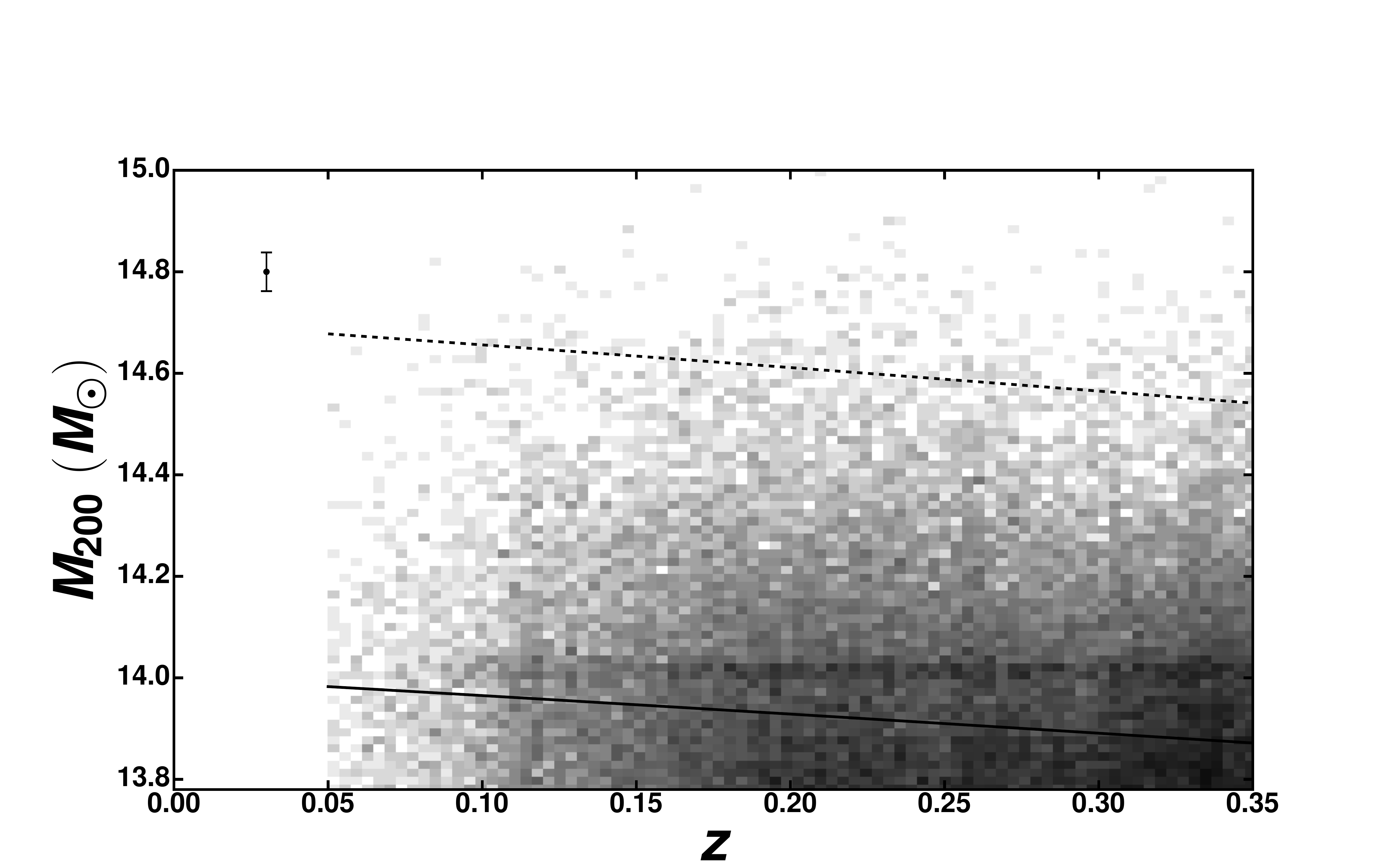}
%
%
%
\caption{Selection of the cluster sample. The grey shaded area represents BCGs in the WHL15 catalogue in clusters with halo masses  $M_{200}/M_\odot > 6.0 \cdot 10^{13}$. 
The black solid line represents the accretion history for a halo with $z=0$ mass $10^{14} M_\odot$, the black dotted line represents the accretion history for a halo with $z=0$ mass $10^{14.7} M_\odot$. 
The black circle with error bars represents the median error on halo mass for all the clusters in the WHL15 catalogue at $0.05 \leq z <0.35$ (prior to applying any cut on redshift and photometry quality). The lines correspond to the boundaries of the 3 cluster sub-samples studied in this work (groups, low to intermediate mass clusters - Virgo-like, and high mass clusters - Coma-like). See text for details.}
\label{fig:plot_accretion_histories_vs_data}
\end{figure}

Objects with a $z=0$ halo mass in the range 
$6 \times 10^{13} < M_{200}/M_\odot< 1.0 \times 10^{14}$ defined our sample of {\itshape{groups}}, while objects with a $z=0$ halo mass in the range 
$1.0 \times 10^{14} \leq M_{200}/M_\odot < 5.0 \times 10^{14}$ defined our sample of Virgo-like, low-to intermediate mass clusters, and objects with a $z=0$ halo mass in the range $M_{200}/M_\odot \geq 5.0 \times 10^{14}$ defined our sample of Coma-like massive clusters. The three sub-samples are shown in Figure \ref{fig:plot_accretion_histories_vs_data}, which also shows the median error on $\log(M_{200}/M_\odot)$ derived from Equation 2 of \cite{Wen_2012}.

Table \ref{table1} presents the detail of the cluster sub-samples after the cuts in optical photometry quality and photometric redshifts described in Section 2 and with the redshift binning introduced in Section 3.2. Since we take into account the accretion history of the haloes, the sub-samples only include the high-redshift progenitors of the clusters that at $z=0$ define the halo mass boundaries. This ensures that the following analysis of BCG properties is robust to progenitor bias.

\begin{table}
  \caption{Details of the sample of brightest cluster galaxies used in the present work}
  \begin{minipage}{12 cm}
  \begin{tabular}{|c|c|c|c|}
    \hline
     \multicolumn{1}{|c}{Redshift}  & \multicolumn{1}{c}{Groups} & \multicolumn{1}{c}{Low-mass Clusters} & \multicolumn{1}{c}{High-mass Clusters} \\
     \hline
     \hline
     $0.05-0.15$ & 1415 & 1882 & 47  \\
     $0.15-0.25$ & 3436 & 4696 & 108 \\
     $0.25-0.35$ & 2281 & 4744 & 97  \\
     \hline
     $0.05-0.35$ & 7132 & 11322 & 252 \\ 
     \hline
  \end{tabular}
\end{minipage}
\label{table1}
\end{table}

\subsection{Optical Colour Distributions} \label{optical_color_distributions}

Optical colours are indicators of the stellar populations in galaxies. Galaxies are clearly observed to have a bimodal colour distribution that separates them into a blue cloud of mostly star-forming systems and a red sequence of passively evolving objects when plotted in a colour-magnitude diagram (see e.g.\ \citealt{Strateva_2001}, \citealt{Baldry_2006}). The red sequence is interpreted as a stellar mass vs metallicity sequence: more massive galaxies are also more metal-rich, while the scatter of the sequence is related to the age and age spread of galaxies (\citealt{Kodama_1997}, \citealt{van_Dokkum_1998}). In general a galaxy can be red because of a high metallicity or a high stellar age, and separating the two effects is not possible using a single colour (\citealt{Worthey_1994}). Thus, combinations of optical and NIR colours must be used in order to disentangle the two effects.

\cite{Eminian_2008} showed that the SDSS $g-r$ rest-frame colour and the UKIDSS\footnote{United Kingdom Infra-red Telescope (UKIRT) Infrared Deep Sky Survey.} $Y-K$ rest-frame colour are respectively sensitive to stellar age and metallicity of galaxies. However, the authors point out that while the $g-r$ rest-frame colour is able to reliably separate galaxies with different ages, going from the youngest to the oldest objects as the colour becomes redder, the dependence of $Y-K$ on metallicity is weak, and only allows one to distinguish galaxies with metallicity lower than solar from those that have metallicity solar and above. For the latter, Figure 11 in \cite{Eminian_2008} shows that there is clearly no dependence between metallicity and $Y-K$ rest-frame colour. BCGs have high metallicities, in general solar or higher (see e.g. \citealt{Thomas_2005}, \citealt{Loubser_2009}, \citealt{Brough_2007}), so the $Y-K$ colour would be of little use in the present study. We therefore only focus on the study of the $g-r$ colour to trace differences in stellar age in our sample of BCGs.

\begin{figure*} 
  \centering 
	\includegraphics[width=0.9\textwidth, trim=0.0cm 0.0cm 0.0cm 0.0cm, clip, page=1]{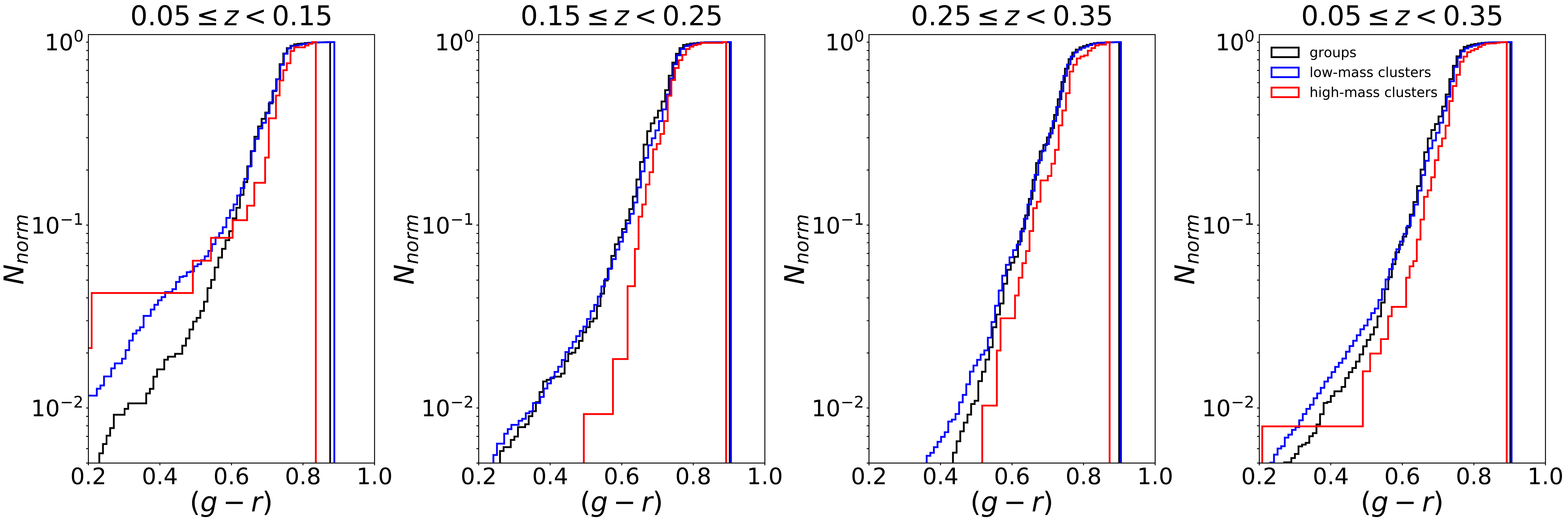}
	\caption{Optical colour distributions of BCGs. Rest-frame $(g-r)$ cumulative colour distributions for the three sub-samples of clusters selected in the WHL15 catalogue in three redshfit bins up to $z=0.35$. The first panel from the right plots the cumulative distributions across the entire redshift range considered in this work. At $0.15 \leq z<0.35$ lower mass clusters and groups have a higher fraction of blue BCGs, suggesting that they host younger central galaxies. Counts in the cumulative histograms are normalised so that their areas are 1. In all the plots black corresponds to groups, blue to low-mass and red to high-mass clusters.}
\label{fig:plot_BCG_color_distributions_Wen_et_al_2015_narrow_range}
\end{figure*}

Following \cite{Eminian_2008}, we estimated the rest-frame $g-r$ colour using the k-corrections provided in the DR12 photometric redshift catalogue of \cite{Beck_2016}. We divided our sample in three redshift bins, namely $0.05 \leq z_{\rm phot} < 0.15$, $0.15 \leq z_{\rm phot} < 0.25$ and $0.25 \leq z_{\rm phot} < 0.35$. Table \ref{table1} details the distribution of the BCGs in each sub-sable and redshift bin, while Figure \ref{fig:plot_BCG_color_distributions_Wen_et_al_2015_narrow_range} shows the $g-r$ rest-frame colour distribution for each cluster sub-sample and redshift bin.

In Fig. \ref{fig:plot_BCG_color_distributions_Wen_et_al_2015_narrow_range} we can see that, at least for BCGs at redshifts $z>0.15$, clusters with low halo masses are more abundant in blue BCGs ($g-r < 0.6$). Tables \ref{table2}, \ref{table3} and  \ref{table4} show the biweight location and scale estimators (\citealt{Beers_1990}) for the rest-frame $(g-r)$ colour distributions in each cluster sub-sample. BCGs in all the sub-samples have consistent values of the biweight locations, implying that the centres of their distributions do not change with cluster mass.

\begin{figure*}
  \centering
	\includegraphics[width=0.9\textwidth, trim=0.0cm 0.0cm 0.0cm 25.0cm, clip, page=1]{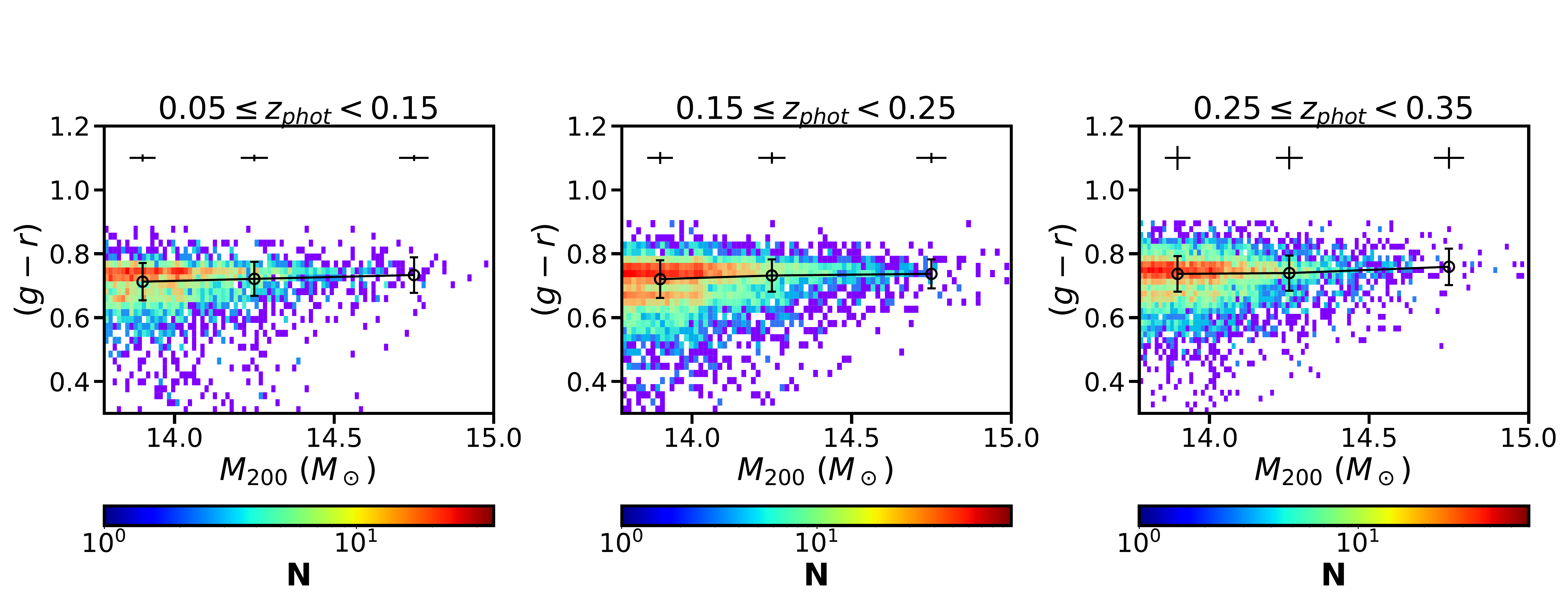}
	\caption{Rest-frame $(g-r)$ colour as a function of cluster dark matter halo mass in the three redshift bins considered in the present work up to $z=0.35$. Error bars on the top of the panels represent the median uncertainties in colour and halo mass in three bins of halo mass ($\log{(M_*/M_\odot)}=13.8-14.0$, $\log{(M_*/M_\odot)}=14.0-14.5$, $\log{(M_*/M_\odot)}=14.5-15.5$). Circles with error bars connected by the solid black line represent the values of the biweight location and scale of the colour distributions in the same halo mass bins. There is no apparent correlation between optical rest-frame colour and halo mass of clusters; however, it can be seen that there appears to be a tail of blue BCGs that extends at low to intermediate cluster masses ($\log(M_{200}/M_\odot) < 14.5$).}
\label{fig:plot_g_r_BCG_color_vs_cluster_mass_kcorr}
\end{figure*}

Figure \ref{fig:plot_g_r_BCG_color_vs_cluster_mass_kcorr} presents the rest-frame $(g-r)$ colour plotted as a function of the cluster halo mass in the three redshift bins introduced above. It can be seen that the rest-frame colour does not correlate with halo mass (a Spearmann correlation test returns a probability $< 1$\% that colour and halo mass are correlated). We see instead that a tail of bluer BCGs appears below $10^{14.5}M_\odot$, the effect being stronger at $z>0.15$. 

A Kolmogorov-Smirnov (KS) and a Mann-Witney U tests return probabilities $<10$\% that the colour distributions of the three sub-samples are consistent among them. The only exception is the U test applied to the colour distributions of the low halo mass and high halo mass clusters at $z = 0.05 - 0.15$, which returns a probability of 12\% that the two samples were drawn from the same population. We note here that the result of the U test is particularly interesting because this statistical test tells one whether a distribution is more skewed towards lower or higher values with respect to another distribution. In other words, the result of this test is telling us that the tail of blue BCGs in lower mass haloes is statistically significant. In Figure \ref{fig:plot_g_r_BCG_color_vs_cluster_mass_kcorr} we also plot the biweight location for the rest-frame colour in three halo mass bins, namely $\log{(M_*/M_\odot)}=13.78-14.0$, $\log{(M_*/M_\odot)}=14.0-14.5$ and $\log{(M_*/M_\odot)}=14.5-15.5$, approximately corresponding to the boundaries of the cluster sub-samples at $z=0$. It can be seen that the values of the biweight location keep constant as a function of mass in all the redshift bins considered. With this estimator being more robust to outliers than the median, this result further supports the notion that there is no correlation between rest-frame $(g-r)$ colour and cluster halo mass, suggesting that the blue BCGs observed at low halo masses constitute a population of outliers in the general trend.

From Figures \ref{fig:plot_BCG_color_distributions_Wen_et_al_2015_narrow_range} and \ref{fig:plot_g_r_BCG_color_vs_cluster_mass_kcorr} we see that BCGs have colours consistent with those of quiescent, red sequence galaxies. However, we note that there are also galaxies with colours as blue as the blue cloud (see e.g.\ \citealt{Hogg_2004}). 
We define blue BCGs as those with $g-r < (g-r)_{\rm median}-2\sigma$, where $(g-r)_{\rm median}$ is the median of the colour distribution. There are in total 1064 blue BCGs, corresponding to 6\% of the sample studied in this work, and the majority of them are located in groups and low halo mass clusters (368 and 683, respectively), with only 13 found in high-mass clusters.

\begin{table}
  \caption{Estimates of the biweight location and scale of rest-frame $g-r$ colour and stellar mass for the {\itshape{groups}} sub-sample.}
  \begin{minipage}{14 cm}
  \begin{tabular}{|c|c|c|c|c|}
    \hline
     \multicolumn{1}{|c}{Redshift} & \multicolumn{1}{c}{$g-r$} & \multicolumn{1}{c}{$\delta (g-r)$} & \multicolumn{1}{c}{$\log(M_*/M_\odot)$} & \multicolumn{1}{c}{$\delta \log(M_*/M_\odot)$} \\
     \hline
     \hline
     $0.05-0.15$ & 0.71 & 0.06 & 11.3 & 0.2 \\
     $0.15-0.25$ & 0.72 & 0.06 & 11.4 & 0.2 \\
     $0.25-0.35$ & 0.74 & 0.06 & 11.4 & 0.2 \\
     \hline
     $0.05-0.35$ & 0.72 & 0.06 &      &     \\ 
     \hline
  \end{tabular}
\end{minipage}
\label{table2}
\end{table}

\begin{table}
  \caption{Estimates of the biweight location and scale of rest-frame $g-r$ colour and stellar mass for the {\itshape{low halo mass clusters}} sub-sample.}
  \begin{minipage}{14 cm}
  \begin{tabular}{|c|c|c|c|c|}
    \hline
    \multicolumn{1}{|c}{Redshift} & \multicolumn{1}{c}{$g-r$} & \multicolumn{1}{c}{$\delta (g-r)$} & \multicolumn{1}{c}{$\log(M_*/M_\odot)$} & \multicolumn{1}{c}{$\delta \log(M_*/M_\odot)$} \\
     \hline
     \hline
     $0.05-0.15$ & 0.72 & 0.06 & 11.4 & 0.3 \\
     $0.15-0.25$ & 0.73 & 0.05 & 11.4 & 0.3 \\
     $0.25-0.35$ & 0.74 & 0.06 & 11.4 & 0.2 \\
     \hline
     $0.05-0.35$ & 0.73 & 0.06 &      &     \\ 
     \hline
  \end{tabular}
\end{minipage}
\label{table3}
\end{table}

\begin{table}
  \caption{Estimates of the biweight location and scale of rest-frame $g-r$ colour and stellar mass for the {\itshape{high halo mass clusters}} sub-sample.}
  \begin{minipage}{14 cm}
  \begin{tabular}{|c|c|c|c|c|}
    \hline
    \multicolumn{1}{|c}{Redshift} & \multicolumn{1}{c}{$g-r$} & \multicolumn{1}{c}{$\delta (g-r)$} & \multicolumn{1}{c}{$\log(M_*/M_\odot)$} & \multicolumn{1}{c}{$\delta \log(M_*/M_\odot)$} \\
    \hline
     \hline
     $0.05-0.15$ & 0.74 & 0.04 & 11.6 & 0.2 \\
     $0.15-0.25$ & 0.74 & 0.05 & 11.6 & 0.3 \\
     $0.25-0.35$ & 0.76 & 0.05 & 11.6 & 0.3 \\
     \hline
     $0.05-0.35$ & 0.74 & 0.05 &      &     \\ 
     \hline
  \end{tabular}
\end{minipage}
\label{table4}
\end{table}

\begin{figure}
  \centering
	\includegraphics[width=0.5\textwidth, trim=0.0cm 0.0cm 0.0cm 0.0cm, clip, page=1]{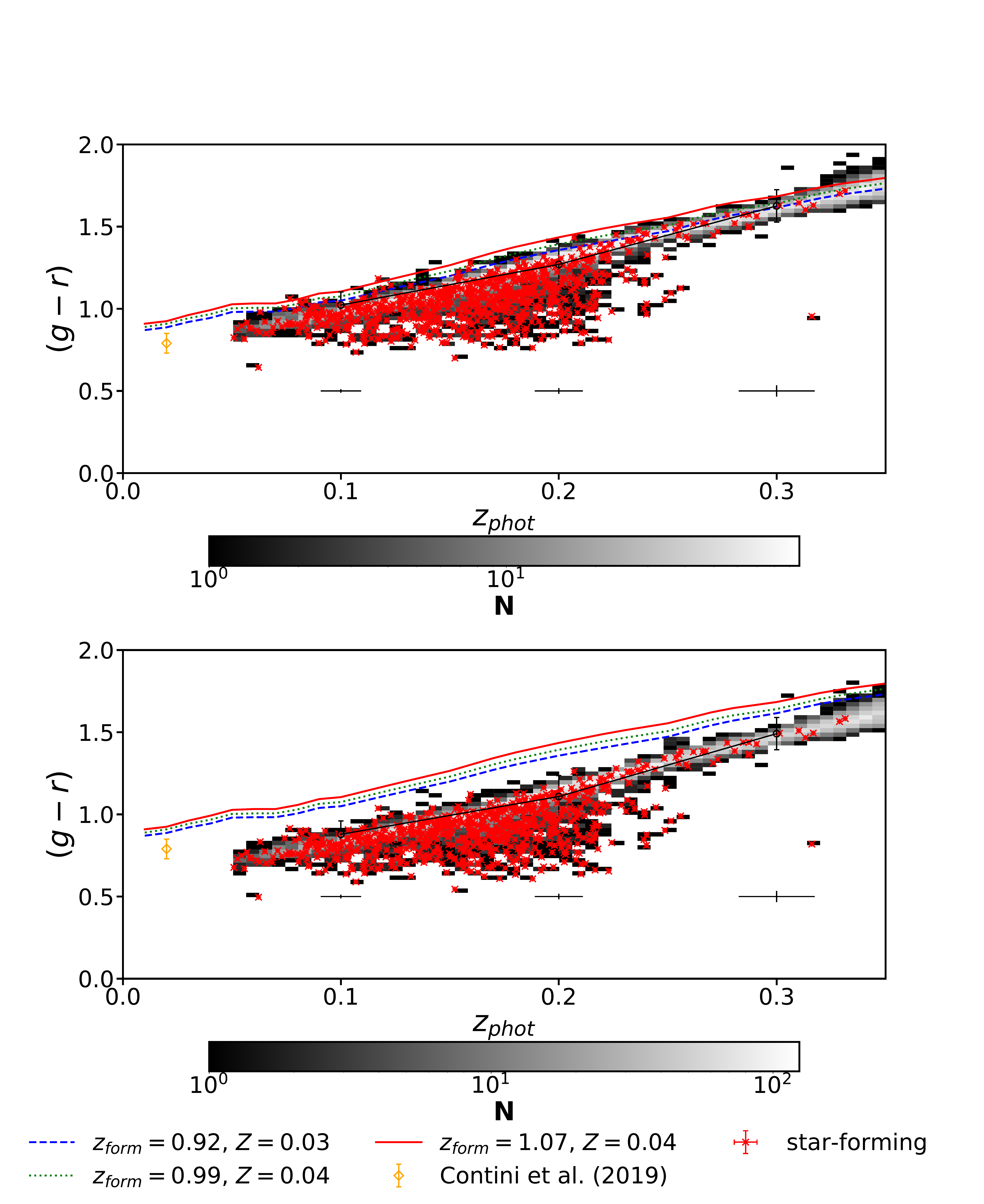}
	\caption{{\itshape{Top panel}}: observer-frame $(g-r)$ colour as a function of cluster redshift. Overplotted are the $(g-r)$ observer-frame colour evolution curves (red, green and blue in order of decreasing stellar mass) predicted by SSP models with formation redshifts ($z_{form}$) and metallicities ($Z$) as from the scaling relations of \protect\cite{Loubser_2009}. The models over-predict the observer-frame colours. The error bars in the bottom of the plot represent the median uncertainties in $(g-r)$ colour and photometric redshift in the three redshift bins considered in this paper. Symbols with error bars connected by the solid black line correspond to the biweight location and scale of the colours in the same redshift bins. {\itshape{Bottom panel}}: the same after applying the correction for missing light in the outskirts of the galaxies.}
\label{fig:plot_BCG_observed_color_vs_models}
\end{figure}

Figure \ref{fig:plot_BCG_observed_color_vs_models} (upper panel) presents the evolution of the observer-frame $(g-r)$ colours in our sample of BCGs. The curves represent the colours derived from the scaling relations for stellar age and metallicity in BCGs at $z=0$ of \cite{Loubser_2009}. Each colour represents a stellar mass at $z=0$ ($10^{11}M_\odot$, blue dashed; $10^{11.5}M_\odot$, green dotted; $10^{12}_\odot$, red solid). 
The curves were obtained using single stellar population (SSP) models with a \cite{Salpeter_1955} initial mass function (IMF) drawn from the 2007 version of the \cite{Bruzual_2003} library\footnote{We use the {\ttfamily{EzGal}} {\ttfamily{python}} library \citep{Mancone_2012} for comparison with stellar population models in this paper}. 
We notice that the colours predicted by all the models are redder than those measured in the BCGs.

Since the model colour tracks are derived assuming age and metallicities of BCGs at $z=0$, the choice of SSP models instead of models with more gradually declining star formation histories or the IMF may have produced such a difference. On the other hand, \cite{von_der_Linden_2007} pointed out that SDSS magnitudes may underestimate the luminosity of extended galaxies such as BCGs because the apertures used miss the light from the outskirts.

We corrected for missing light using an approach similar to \cite{DSouza_2015}. We split each of the $z_{\rm phot}$ bins in which our sample was divided in two $r$-band Petrosian magnitude bins, $r<17.0$ and $r \geq 17.0$. We then cut 200$\times$200 pixel postage-stamp images\footnote{This corresponds to 79.2$''$, which is equivalent to a projected physical size of $\sim 75$ kpc at $z=0.05$ and $\sim 390$ kpc at $z=0.35$.} of the galaxies from the parent SDSS $g$- and $r$- band frames and for each redshift-magnitude bin we derived the median images in the two bands.

We then used the {\ttfamily{iraf}}\footnote{Image Reduction and Analysis Facility. IRAF is distributed by the National Optical Astronomy Observatory, which is operated by the Association of Universities for Research in Astronomy (AURA) under a cooperative agreement with the National Science Foundation.} task {\ttfamily{ellipse}} to estimate the total flux at different isophotal positions in the median galaxy images and, in particular, we considered the magnitude at 4 half-light radii as the total magnitude of the median galaxies.

We determined the differences between these magnitudes and the median model magnitudes and used this as the correction term ($\Delta g$ and $\Delta r$) for model magnitudes in the $g$ and $r$ bands in each redshift and magnitude bin. The corrections vary with redshift, magnitude and photometric bands and range between 0.002 and 0.5 mag. In particular, we note that for bright galaxies the corrections decrease with redshift in both the $g$ and $r$ bands, while for faint galaxies they increase with redshift. We also note that $\Delta g > \Delta r$. The values of $\Delta g$ and $\Delta r$ are reported in Table \ref{table5}.

We note, in the bootom panel of Figure \ref{fig:plot_BCG_observed_color_vs_models}, that even with the correction for missing light, the observed $g-r$ colours of our BCGs are bluer than the predictions of the models. Furthermore, since $\Delta g > \Delta r$, the colours are bluer than those derived from model magnitudes. 
From Fig. \ref{fig:plot_BCG_observed_color_vs_models} we can see that star-forming BCGs (red symbols, see next sections) populate the entire diagram with the exception of the reddest colours at all redshifts, so the application of the photometric corrections does not depend on the star-formation activity.

We plotted as an orange diamond in Figure \ref{fig:plot_BCG_observed_color_vs_models} the median and the $1\sigma$ width of the BCG colour distribution predicted by the theoretical models of \cite{Contini_2019}. These semi-analytic models predict $g-r$ colours that are bluer than those inferred from stellar populations in $z=0$ BCGs and are consistent with the colours observed in our sample.

We attribute the disagreement between the SSP colours and the predictions of the \cite{Contini_2019} models to the different areas considered in the two works to study stellar populations. \cite{Loubser_2009} limited indeed themselves to within $\sim 1/8$ half-light radii from the centre of the galaxies, while  \cite{Contini_2019} considered a larger extent (out to $\sim 10 \,$ kpc). As discussed in \cite{Contini_2019} and \cite{Loubser_2012}, brightest cluster galaxies may have negative metallicity gradients, becoming gradually metal poorer as one moves away from the centres.

Metallicity gradients produce colour gradients, and galaxies become bluer at large distances from the centre as larger fractions of the metal-poor stellar populations are included in the photometric apertures. \cite{Contini_2019} predict a specially low metallicity for BCGs, which at 3 kpc is lower than solar. The presence of metallicity gradients and the large apertures considered in this work to derive galaxy colours can explain the bluer colours that we observe in our sample of BCGs with respect to those predicted assuming the \cite{Loubser_2009} scaling relations.

In the following we shall use the $g-r$ colours obtained from the model magnitudes without applying the photometric correction for the missing light. Since we are focused on the investigation of colour differences in different sub-sets of BCGs, our conclusions are not affected by the loss of light in their outskirts. Interestingly, the trend of the observed BCG colours with redshift appears consistent with passive evolution.

\begin{table}
  \caption{Photometric corrections for missing light in the wings of the light profiles of BCGs. All quantities are AB magnitudes.}
  \begin{minipage}{14 cm}
  \begin{tabular}{|c|c|c|c|}
    \hline
    \multicolumn{1}{|c}{Redshift} & \multicolumn{1}{c}{$r_{petro}$} & \multicolumn{1}{c}{$\Delta g$} & \multicolumn{1}{c}{$\Delta r$} \\ 
    \hline
    \hline
     $0.05-0.15$ & $< 17$    & $0.4$   & $0.3$    \\
     $0.05-0.15$ & $\geq 17$ & $0.12$  & $0.03$   \\
     $0.15-0.25$ & $< 17$    & $0.4$   & $0.2$     \\
     $0.15-0.25$ & $\geq 17$ & $0.3$   & $0.17$     \\
     $0.25-0.35$ & $< 17$    & $0.06$  & $0.0017$   \\
     $0.25-0.35$ & $\geq 17$ & $0.5$   & $0.4$      \\
    \hline
  \end{tabular}
\end{minipage}
\label{table5}
\end{table}

\subsection{Infra-Red Colours and Detection of Star-forming BCGs} \label{star_forming_BCG}

We use IR colours to search for BCGs with ongoing star formation activity. The $W1$ and $W2$ filters trace the continuum emission from low-mass evolved stars, with low dust extinction, while the $W3$ filter, spanning the region of the electromagnetic spectrum dominated by polycyclic aromatic hydrocarbons (PAH) and $\rm[Ne II]$ emission, is sensitive to the interstellar medium. Finally, the $W4$ band spans the spectral region where there is the light emitted by dust as a consequence of the heating of the grains by star formation and/or AGN activity (\citealt{Jarrett_2011}, \citealt{Popescu_2011}, \citealt{Cluver_2014}, \citealt{Leja_2018}).


As shown in \cite{Wright_2010}, \cite{Jarrett_2011} and \cite{Jarrett_2017}, one can use the $W1-W2$ vs $W2-W3$ colour-colour diagram to separate a population of galaxies into quiescent, star-forming and AGN hosts: quiescent galaxies, approximately corresponding to morphologically elliptical and S0 galaxies, occupy the regions with $(W1-W2)<0.8$ and $(W2-W3) < 1.5$; star-forming galaxies, approximately corresponding to morphologically spiral and  irregular galaxies, occupy the region with $(W1-W2)<0.8$ and $(W2-W3) \geq 1.5$; AGN hosts and ULIRGS reside in the region with $(W1-W2)\geq0.8$. 

\begin{figure*}
  \centering
	\includegraphics[width=0.9\textwidth, trim=0.0cm 0.0cm 0.0cm 0.0cm, clip]{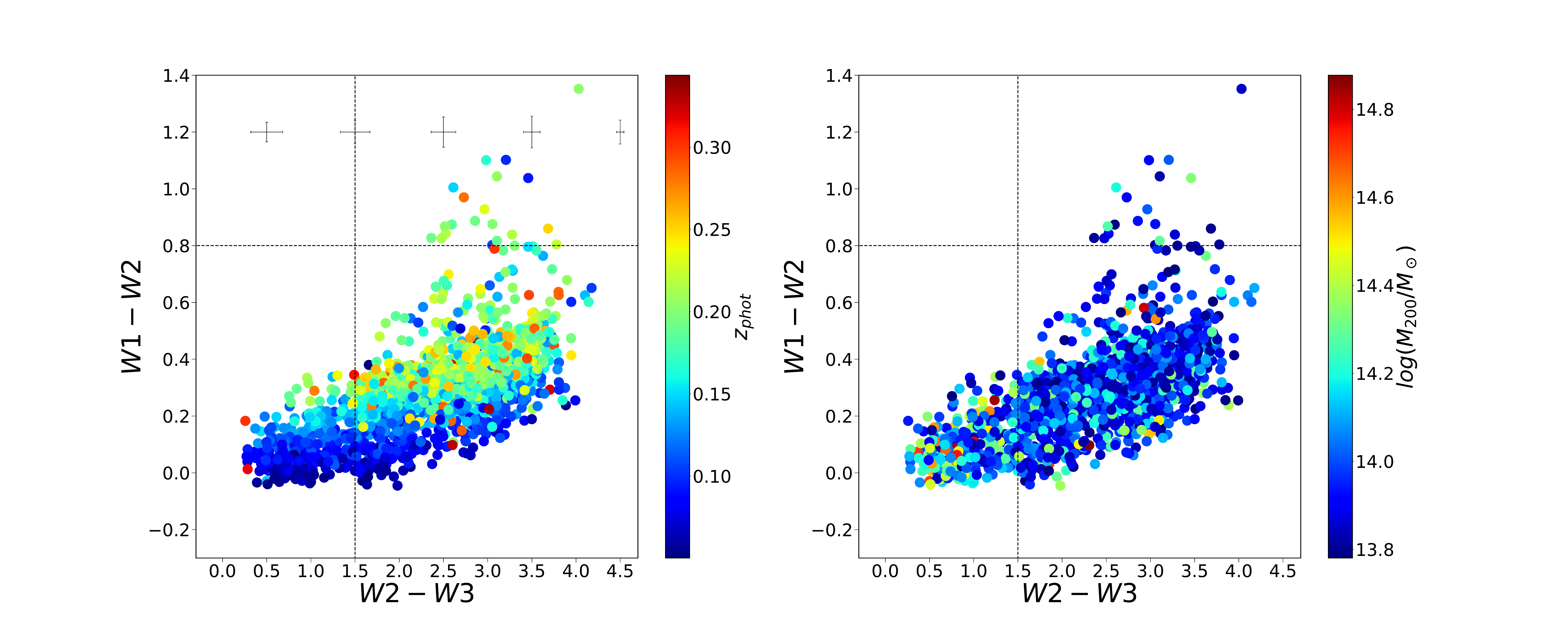}
	\caption{The WISE $W1-W2$ vs $W2-W3$ colour-colour diagram. On the left circles are colour coded according to the photometric 
redshifts of the galaxies and on the right they are colour coded according to the halo masses of the clusters. Dashed lines show the subdivision of the diagram into different classes of quiescent galaxies, star-forming galaxies and AGN hosts as explained in the text. The black crosses represent the median errors on colours. The objects represented in this plot are those selected in $W1$, $W2$ and $W3$, which correspond to $\sim$10\% of the sample at $0.05 \leq z < 0.35$ (See Section \ref{IR_data})}.
\label{fig:colour-colour_plots}
\end{figure*}

The colour-colour diagram for the BCGs in our sample with detections in  W1, W2 and W3 (see Fig. \ref{fig:colour-colour_plots}), shows that the majority of the galaxies are in the quiescent and star forming regions ($\displaystyle f_{\rm Q} = 0.174_{-0.008}^{+0.009}$ and $\displaystyle f_{\rm SF} = 0.815_{-0.009}^{+0.009}$, respectively). Only a minority of the galaxies are found in the AGN region of the diagram ($\displaystyle f_{\rm AGN} = 0.011_{-0.002}^{+0.003}$). As a comparison, in \cite{Fraser_McKelvie_2014}, who used the same regions of the WISE colour-colour space to distinguish among quiescent, star-forming and AGN host galaxies at $z<0.1$, only 3\% of their BCG sample consists of star-forming galaxies, while the remaining 97\% is made of quiescent systems.

As shown in Appendix A, the sample of BCGs that we use to detect star formation in these galaxies is essentially $W3$-selected. This results in favouring the occurrence of star-forming BCGs, which are strong emitters at $12 \mu$m, over the occurrence of quiescent objects. This explains the fact that our sub-sample of WISE-detected BCGs is dominated by star-forming objects.

\citet{Jarrett_2017} in the GAMA G12 sample (\citealt{Driver_2011}) show that the galaxies in the star-forming region (intermediate and star-forming discs in their nomenclature) are systems that reside in the main sequence and in the transition zone between the main sequence and the quiescent locus in the star formation rate (SFR) vs stellar mass ($M_*$) diagram. We note that the star-forming region in the WISE colour-colour diagram includes galaxies with $SFR$ as low as $0.1 M_\odot \cdot$yr$^{-1}$, which are below the main sequence (\citealt{Speagle_2014}). 
Galaxies that are therefore star-forming according to the colour-colour diagram may be considered non star-forming according to their position with respect to the main sequence. We take this as a caveat for the discussion of the results, while we address the study of star formation in BCGs in Orellana et al. (in prep.).

We observe here that BCGs with robust detections in $W1$, $W2$ and $W3$ are only 10\% of the sample, and those with detections in $W4$ are even less (0.8\%), as a result of the broad PSF and low $S/N$ at $22 \mu m$. Although they dominate our sample of simultaneous detections in $W1$, $W2$ and $W3$, BCGs with ongoing star formation represent a minority in the overall sample of $z<0.35$ clusters that we are studying in this paper. Star-forming BCGs are therefore rare in the nearby Universe, in agreement with other works at similar redshifts (e.g.\ \citealt{Fraser_McKelvie_2014}, \citealt{Oliva_Altamirano_2014}).

\begin{figure*}
  \centering
	\includegraphics[width=0.9\textwidth, trim=0.0cm 0.0cm 0.0cm 0.0cm, clip, page=1]{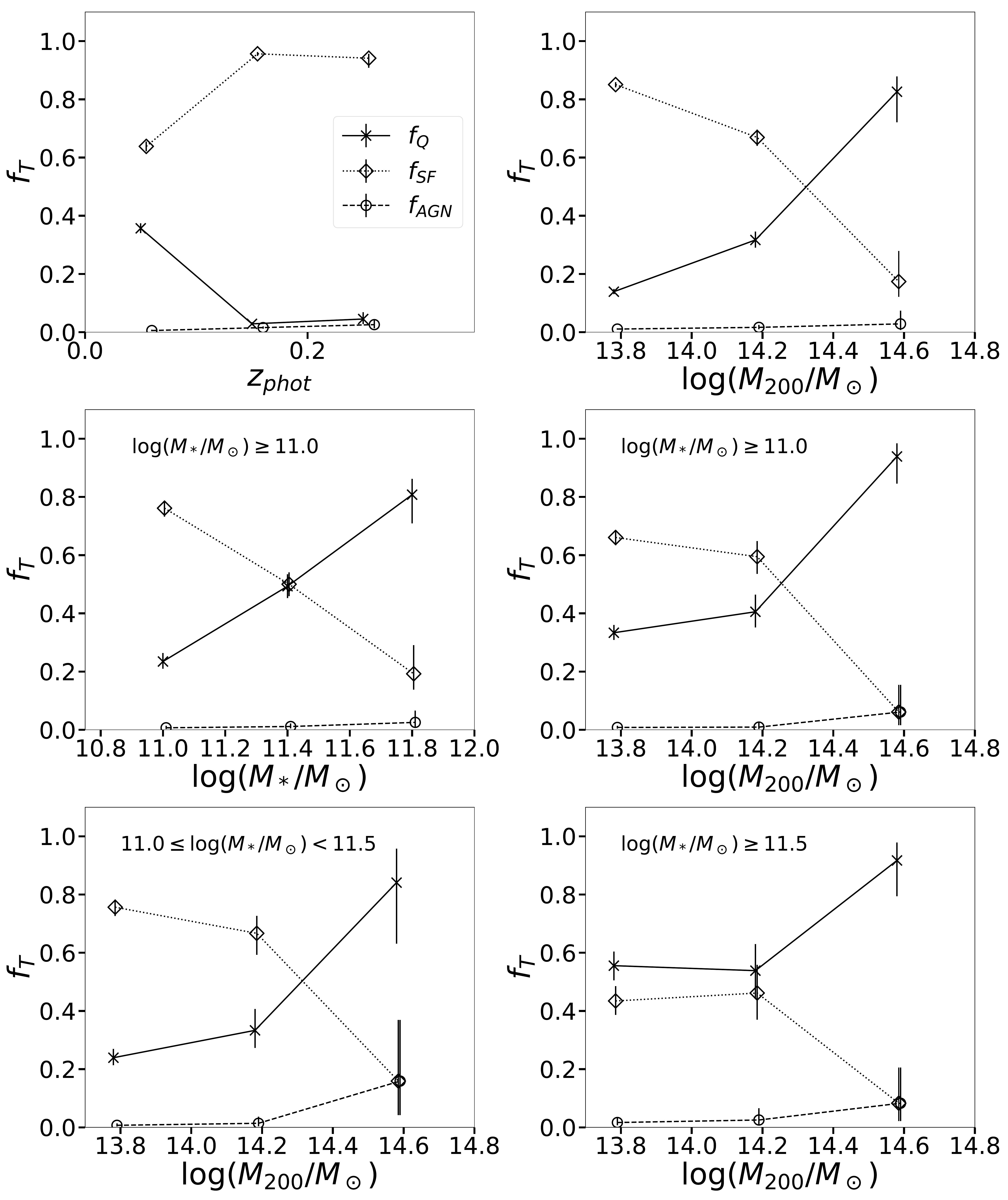}
	\caption{{\itshape{Top left}}: fraction $f_{\rm T}$ of galaxy types as a function of photometric redshift. {\itshape{Top right}}: fraction of galaxy types as a function of cluster halo mass. Bins are the same shown in Tables \ref{table5} and \ref{table6}.   {\itshape{Middle left}}: fraction of galaxy types as a function of BCG stellar mass. Bins are the same shown in Table \ref{table10}. {\itshape{Middle right}}: fraction of galaxy types as a function of cluster halo mass for galaxies with stellar masses $M_* \geq 10^{11} M_\odot$. {\itshape{Bottom left}}: fraction of galaxy types as a function of cluster halo mass for galaxies with stellar masses $10^{11}\mbox{ }M_\odot \leq M_* < 10^{11.5}\mbox{ }M_\odot$. {\itshape{Bottom right}}: fraction of galaxy types as a function of cluster halo mass for galaxies with stellar masses $M_* \geq 10^{11.5}\mbox{ }M_\odot$. The bins in the bottom panels and in the middle right panel are as in Table \ref{table6}. The fraction of star-forming BCGs $f_{SF}$ grows with redshift and decreases with halo mass and stellar mass.}
\label{fig:plot_WISE_type_fractions}
\end{figure*}

In Figure \ref{fig:colour-colour_plots} we can see that star-forming BCG are more frequent at high redshifts and in lower mass haloes. Tables \ref{table6} and \ref{table7} show the fractions of galaxy types as a function of photometric redshift and halo mass in the sub-sample of BCGs with simultaneous detections in $W1$, $W2$ and $W3$. These fractions are plotted in the top panels of Figure \ref{fig:plot_WISE_type_fractions}. We chose bins of 0.4 dex in halo mass, which is 4 times larger than the median halo mass error. It can be seen that the fraction of star-forming galaxies increases from 60\% to 95\% from $z \sim 0.1$ to $z \sim 0.2$ and then stabilises itself at $\sim 95$\%. At the same time we also note that the fraction of star-forming galaxies drops at $\log{(M_{200}/M_*)} > 14.2$ from $\sim 70$\% to $\sim 20$\%.

In the next two sections we will investigate the trends of the quiescent and star-forming fractions with the stellar mass of the BCGs and the effects that this quantity has in determining the star-formation activity of those galaxies.

\begin{table}
  \caption{Fractions of galaxy types as a function of photometric redshift for the BCGs in the sub-sample detected in W1, W2 and W3. The last column in the table indicates the number of galaxies $N_g$ in each bin.}
  \begin{minipage}{14 cm}
  \begin{tabular}{|c|c|c|c|c|}
    \hline
    \multicolumn{1}{|c}{Redshift} & \multicolumn{1}{c}{$f_{\rm q}$} & \multicolumn{1}{c}{$f_{\rm SF}$} & \multicolumn{1}{c}{$f_{AGN}$} & \multicolumn{1}{c}{$N_g$}\\
    \hline
     \hline
     $0.05-0.15$ & $0.357_{-0.016}^{+0.017}$ & $0.638_{-0.017}^{+0.016}$ & $0.006_{-0.002}^{+0.003}$ & $824$ \\
     $0.15-0.25$ & $0.029_{-0.005}^{+0.006}$ & $0.956_{-0.007}^{+0.006}$ & $0.016_{-0.004}^{+0.004}$ & $931$ \\
     $0.25-0.35$ & $0.05_{-0.02}^{+0.02}$ & $0.94_{-0.03}^{+0.02}$ & $0.026_{-0.013}^{+0.019}$ & $102$ \\
     \hline
  \end{tabular}
\end{minipage}
\label{table6}
\end{table}

\begin{table}
  \caption{Fractions of galaxy types as a function of cluster halo mass for the BCGs in the sub-sample detected in $W1$, $W2$ and $W3$. The last column in the table indicates the number of galaxies $N_g$ in each bin.}
  \begin{minipage}{14 cm}
  \begin{tabular}{|c|c|c|c|c|}
    \hline
    \multicolumn{1}{|c}{$\log{(M_{200}/M_\odot)}$} & \multicolumn{1}{c}{$f_{\rm q}$} & \multicolumn{1}{c}{$f_{\rm SF}$} & \multicolumn{1}{c}{$f_{AGN}$} & \multicolumn{1}{c}{$N_g$}\\
    \hline
     \hline
    $13.78-14.18$ & $0.139_{-0.008}^{+0.009}$ & $0.851_{-0.010}^{+0.009}$ & $0.011_{-0.002}^{+0.003}$ & $1553$ \\
    $14.18-14.58$ & $0.32_{-0.03}^{+0.03}$ & $0.67_{-0.03}^{+0.03}$ & $0.017_{-0.006}^{+0.009}$ & $281$ \\
    $14.58-14.98$ & $0.83_{-0.11}^{+0.05}$ & $0.17_{-0.05}^{+0.11}$ & $0.03_{-0.02}^{+0.05}$ & $23$ \\
  \hline
  \end{tabular}
\end{minipage}
\label{table7}
\end{table}

\begin{table}
  \caption{Fractions of galaxy types as a function of cluster halo mass for the BCGs with stellar masses $\log{M_*/M_\odot} = 11.0-11.5$ (upper rows), $\log{M_*/M_\odot} \geq 11.5$ (central rows) and $\log{M_*/M_\odot} \geq 11.0$ (bottom rows) in the sub-sample detected in $W1$, $W2$ and $W3$. The last column in the table indicates the number of galaxies $N_g$ in each bin.}
  \begin{minipage}{14 cm}
  \begin{tabular}{|c|c|c|c|c|}
    \hline
    \multicolumn{1}{|c}{$\log{(M_{200}/M_\odot)}$} & \multicolumn{1}{c}{$f_{\rm q}$} & \multicolumn{1}{c}{$f_{\rm SF}$} & \multicolumn{1}{c}{$f_{AGN}$} & \multicolumn{1}{c}{$N_g$}\\
    \hline
     \hline
    $13.78-14.18$ & $0.24_{-0.03}^{+0.03}$ & $0.76_{-0.03}^{+0.03}$ & $0.007_{-0.004}^{+0.007}$ & $234$ \\
      & $0.56_{-0.05}^{+0.05}$ & $0.43_{-0.05}^{+0.05}$ & $0.017_{-0.010}^{+0.016}$ & $99$ \\
      & $0.33_{-0.02}^{+0.03}$ & $0.66_{-0.03}^{+0.02}$ & $0.008_{-0.004}^{+0.006}$ & $333$ \\
$14.18-14.58$ & $0.33_{-0.06}^{+0.07}$ & $0.67_{-0.07}^{+0.06}$ & $0.01_{-0.01}^{+0.02}$ & $48$ \\
      & $0.54_{-0.10}^{+0.09}$ & $0.46_{-0.09}^{+0.10}$ & $0.03_{-0.02}^{+0.04}$ & $26$ \\
      & $0.41_{-0.05}^{+0.06}$ & $0.59_{-0.06}^{+0.05}$ & $0.009_{-0.007}^{+0.015}$ & $74$ \\
$14.58-14.98$ & $0.8_{-0.2}^{+0.1}$ & $0.2_{-0.1}^{+0.2}$ & $0.1_{-0.1}^{+0.2}$ & $3$ \\
      & $0.92_{-0.12}^{+0.06}$ & $0.08_{-0.06}^{+0.12}$ & $0.08_{-0.06}^{+0.12}$ & $7.0$ \\
      & $0.94_{-0.09}^{+0.05}$ & $0.06_{-0.05}^{+0.09}$ & $0.06_{-0.05}^{+0.09}$ & $10$ \\
 \hline
  \end{tabular}
\end{minipage}
\label{table8}
\end{table}

\begin{table}
  \caption{Fractions of galaxy types as a function of BCG stellar mass in the sub-sample detected in $W1$, $W2$ and $W3$. The last column in the table indicates the number of galaxies $N_g$ in each bin.}
  \begin{minipage}{14 cm}
  \begin{tabular}{|c|c|c|c|c|}
    \hline
    \multicolumn{1}{|c}{$\log{(M_*/M_\odot)}$} & \multicolumn{1}{c}{$f_{\rm q}$} & \multicolumn{1}{c}{$f_{\rm SF}$} & \multicolumn{1}{c}{$f_{AGN}$} & \multicolumn{1}{c}{$N_g$}\\
    \hline
     \hline
    $11.0-11.4$ & $0.23_{-0.02}^{+0.03}$ & $0.76_{-0.03}^{+0.03}$ & $0.007_{-0.004}^{+0.007}$ & $243$ \\
    $11.4-11.8$ & $0.49_{-0.04}^{+0.04}$ & $0.50_{-0.04}^{+0.04}$ & $0.011_{-0.006}^{+0.011}$ & $148$ \\
    $11.8-12.2$ & $0.81_{-0.10}^{+0.05}$ & $0.19_{-0.05}^{+0.10}$ & $0.03_{-0.02}^{+0.04}$ & $26$ \\
 \hline
  \end{tabular}
\end{minipage}
\label{table9}
\end{table}

\begin{table}
  \caption{Fractions of galaxy types as a function of BCG stellar mass in the sub-sample detected in $W1$, $W2$ and $W3$ and split according to the halo mass of the clusters. By low and high halo mass clusters we here mean systems that have masses lower and higher than the median halo mass of the sub-sample, respectively. Values for the low (high) halo mass clusters are reported in the top (bottom) rows. The last column in the table indicates the number of galaxies $N_g$ in each bin.}
  \begin{minipage}{14 cm}
  \begin{tabular}{|c|c|c|c|c|}
    \hline
    \multicolumn{1}{|c}{$\log{(M_*/M_\odot)}$} & \multicolumn{1}{c}{$f_{\rm q}$} & \multicolumn{1}{c}{$f_{\rm SF}$} & \multicolumn{1}{c}{$f_{AGN}$} & \multicolumn{1}{c}{$N_g$}\\
    \hline
     \hline
     $11.0-11.4$ & $0.19_{-0.03}^{+0.04}$ & $0.80_{-0.04}^{+0.03}$ & $0.013_{-0.007}^{+0.012}$ & $132$ \\
       & $0.29_{-0.04}^{+0.05}$ & $0.71_{-0.05}^{+0.04}$ & $0.006_{-0.005}^{+0.010}$ & $111$ \\
     $11.4-11.8$ & $0.5_{-0.06}^{+0.06}$ & $0.49_{-0.06}^{+0.06}$ & $0.02_{-0.01}^{+0.02}$ & $68$ \\
       & $0.49_{-0.05}^{+0.06}$ & $0.51_{-0.06}^{+0.05}$ & $0.009_{-0.006}^{+0.014}$ & $80$ \\
     $11.8-12.2$ & $0.7_{-0.2}^{+0.1}$ & $0.3_{-0.1}^{+0.2}$ & $0.08_{-0.06}^{+0.12}$ & $7$ \\
       & $0.84_{-0.12}^{+0.05}$ & $0.16_{-0.05}^{+0.12}$ & $0.03_{-0.03}^{+0.05}$ & $19$ \\
 \hline
  \end{tabular}
\end{minipage}
\label{table10}
\end{table}

\subsection{Stellar Masses} \label{section_stellar_masses_vs_z}

We use stellar masses from the \cite{Maraston_2013} catalogue. These authors used a SED fitting technique to estimate stellar masses and stellar population parameters such as SFR, stellar age and metallicity. In particular, they fitted the SDSS DR10 \citep{Ahn_SDSS_DR10_2014} photometry to SED templates using an algorithm based on the minimization of $\chi^2$ implemented in the {\ttfamily{hyperz}} software \citep{Bolzonella_2000}. 
The authors used a passively evolving SED template with a two component metallicity (97\% solar and 3\% 0.05 solar) and a set of star-forming template SED with exponentially declining (with e-folding time $\tau = 0.1, 0.3, 1.0$ Gyr) and truncated star formation histories. The latter were parametrised as constant star formation histories that lasted for a finite time ($0.1, 0.3, 1.0$ Gyr, truncation time) and then dropped to 0. The templates were drawn from the \cite{Maraston_2009} library, and the stellar masses obtained with the passively-evolving and star-forming sets of templates are stored in the catalogues {\ttfamily{stellarMassPassivePort}} and {\ttfamily{stellarMassStarformingPort}}, respectively\footnote{The catalogues can be downloaded from the SDSS repository at \url{https://skyserver.sdss.org/CasJobs/}.}

The authors assumed a \cite{Kroupa_2001} IMF and a flat $\Lambda CDM$ cosmology with $\Omega_M = 0.25$, $\Omega_\Lambda = 0.75$ and $H_0=73.0$ km~s$^{-1}$~Mpc$^{-1}$ (Wilkinson Microwave Anisotropy Probe, WMAP, 1). 
The derivation of stellar masses is discussed in \cite{Maraston_2013}, and we refer the reader to that paper for the details. 
We matched the WHL15 BCG catalogue with the {\ttfamily{stellarMassStarformingPort}} using a search radius of $\sim0.02''$ (the same adopted in the query of the SDSS photometric and photometric redshift databases, see Section \ref{optical_data}).

\begin{figure*}
  \centering
	\includegraphics[width=\textwidth, trim=0.0cm 0.0cm 0.0cm 0.0cm, clip, page=2]{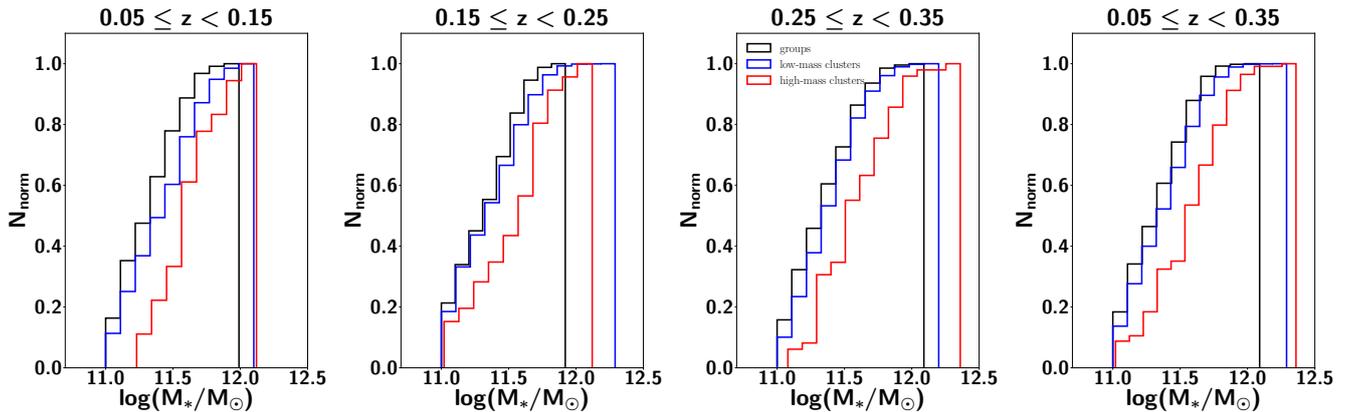}
	\caption{Cumulative distributions of stellar mass in the three sub-samples of BCGs in the present work in the three redshift bins adopted up to $z=0.35$. The first panel from the right shows the distributions across the entire redshift range studied. Lower mass clusters and groups are more abundant in BCGs with low stellar masses. Counts in the cumulative histograms are normalised so that their areas are 1.}
\label{fig:plot_BCG_stellar_mass_Portsmouth_distributions_Wen_et_al_2015}
\end{figure*}

In studying the stellar masses in our sample of BCGs we added to the selection in photometric redshift and $(g-r)$ colour described in \S \ref{optical_color_distributions} the requirement that the fractional uncertainties on stellar mass were lower than 10\%. 
We defined the uncertainty on stellar mass as half the difference between the $1\sigma$ maximum and minimum stellar masses obtained in the range  $\chi^2_{minimum}+1$ ({\ttfamily{maxLogMass}} and {\ttfamily{minLogMass}} in the stellar mass catalogue), where $\chi^2_{minimum}$ is the minimum $\chi^2$ in the fit.

In order to build a stellar mass complete sample, we derived the stellar mass of a passively evolving single stellar population SED template with $r=21.0$ mag, the magnitude of the faintest BCG in the sample. The template was drawn from the 2007 version of the \cite{Bruzual_2003} stellar population library, it has formation redshift $z=5.0$, \cite{Salpeter_1955} IMF and solar metallicity. This selection ensured that the sub-sample of BCGs with stellar masses was not affected by incompleteness at the highest redshift considered in this work. After applying our cuts in photometry quality and photometric redshift reliability, we end up with a stellar mass complete sample of 6708 BCGs.

\begin{figure}
\centering
\includegraphics[width=0.5\textwidth, trim=0.0cm 35.0cm 0.0cm 0.0cm, clip, page=1]{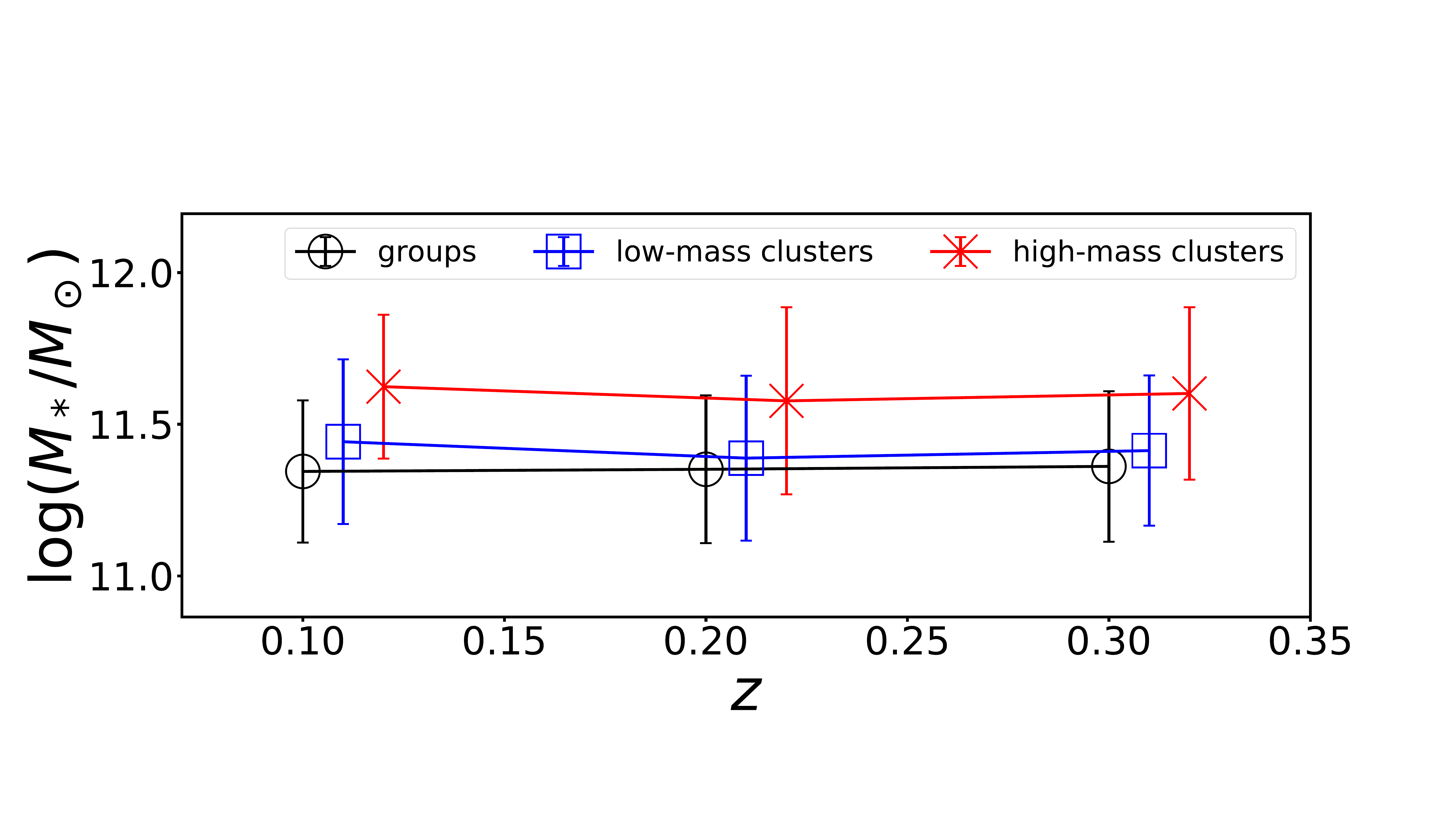}
\caption{Biweight location and scale of BCG stellar mass in the three subsamples of clusters studied in the present work as a function of redshift. BCGs have similar stellar masses in all the subsamples, although lower mass clusters and groups tend to have lower stellar masses. We do not observe significant stellar mass growth at $z<0.35$ in any of the sub-samples of the clusters considered here.}
\label{fig:plot_BCG_stellar_pop_Portsmouth_biweight_estimators_Wen_et_al_2015}
\end{figure}

The cumulative distributions of the stellar masses in the three sub-samples of BCGs studied in this work are shown in Fig. \ref{fig:plot_BCG_stellar_mass_Portsmouth_distributions_Wen_et_al_2015}, while Fig. \ref{fig:plot_BCG_stellar_pop_Portsmouth_biweight_estimators_Wen_et_al_2015} shows the biweight location and scale of the stellar mass distributions in each sub-sample in the three bins of redshifts considered in this paper. The values of the biweight location and scale of the stellar mass distributions are reported in Tables \ref{table2}, \ref{table3} and \ref{table4}.

\begin{figure}
  \centering
	\includegraphics[width=0.5\textwidth, trim=0.0cm 8.0cm 0.0cm 0.0cm, clip, page=1]{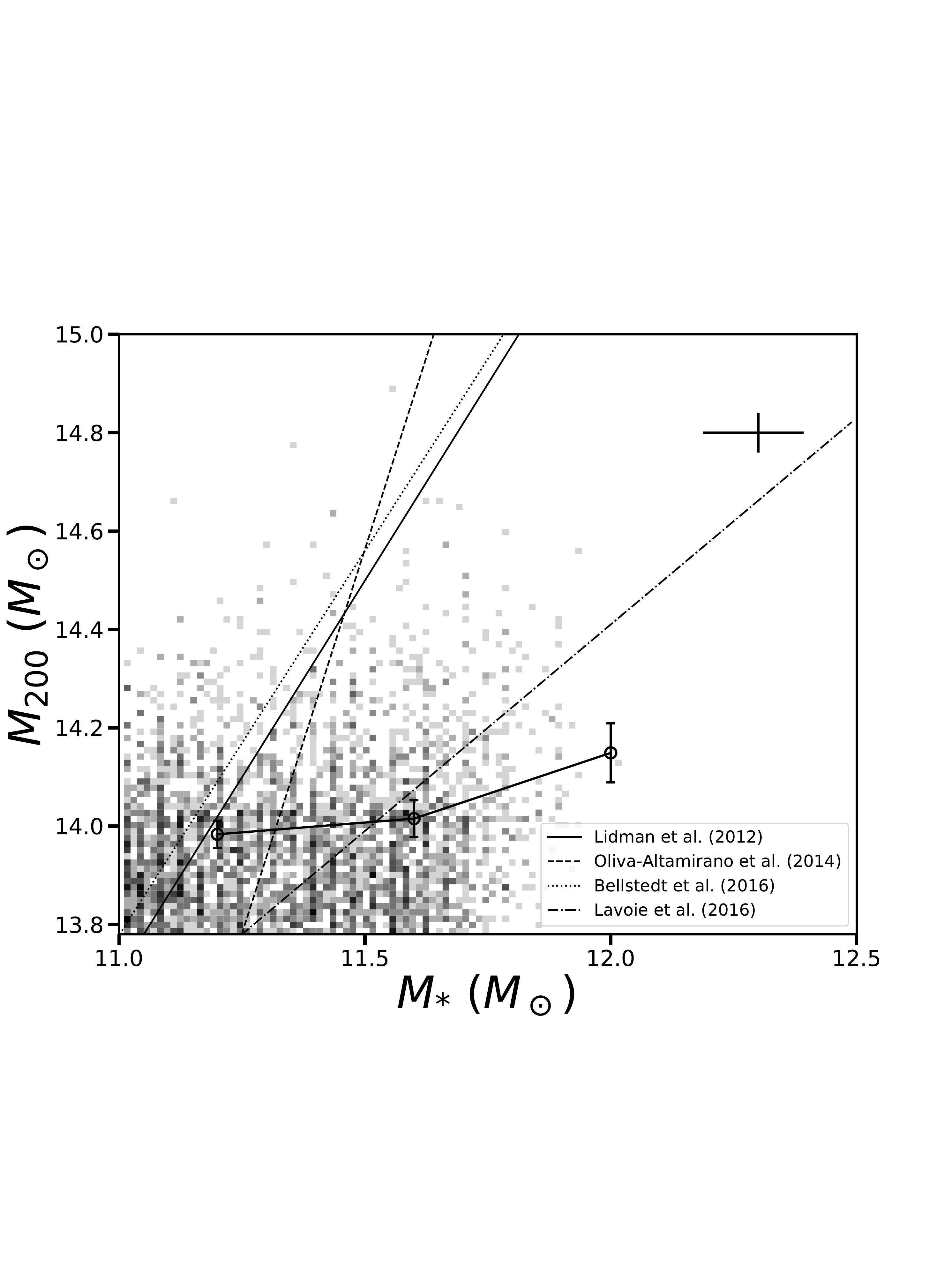}
	\caption{The halo mass of the clusters in the sub-sample with reliable stellar masses at $0.05 \leq z < 0.35$ plotted as a function of the BCG stellar mass. The lines represent recent literature results for the $M_{200}$ vs $M_*$ scaling relation. Circles with error bars represent the median and $1\sigma$ dispersion of $\log(M_{200})$ in bins of $\log(M_{*})$. The flatter trend of the the median values may be a result of the rigid limits that we impose on cluster and BCG stellar mass in this work.}
\label{fig:plot_M200_vs_Mstar}
\end{figure}

We see that higher-mass clusters tend to host BCGs with higher stellar masses and that the values of the biweight location of the stellar mass distributions in the redshift range that we consider are all consistent among them. The correlation between stellar mass of BCGs and the mass of their host halo is well known in the literature (\citealt{Brough_2008}, \citealt{Lidman_2012}, \citealt{Oliva_Altamirano_2014}, \citealt{Bellstedt_2016}). In Fig. \ref{fig:plot_M200_vs_Mstar} we plot the cluster halo mass as a function of stellar mass for the stellar-mass limited sample of BCGs together with the scaling relations presented in some recent works. It can be seen that the median values of $M_{200}$ increase with $M_*$ the increase is shallower than the scaling relations presented in the literature.

Our results also tell us that BCGs do not evolve significantly in stellar mass at $z<0.35$ and that this lack of evolution happens at all halo masses. 
The latter result, which is in agreement with other works that focused on BCGs at similar redshifts (e.g.,\ \citealt{Oliva_Altamirano_2014}) is in tension with theoretical models, which predict that BCGs should grow by $\sim 50$\% in stellar mass at $z<0.5$ \citep{De_Lucia_and_Blaizot_2007}. 
This result suggests that BCGs had already completed their mass assembly at $z \sim 0.4$. We will discuss the implications of these results in Section 4.5.

\section{Discussion}

BCGs are the most massive galaxies in the Universe and the study of their properties is crucial to understand the formation and evolution of galaxies in general. Since BCGs reside in clusters, they are rare and, therefore, a thorough picture of their physical properties can only be achieved with large samples as the one considered in this paper. 
The present section discusses and tries to put together in a coherent picture the results obtained from the analysis of the WHL15 BCG sample presented in Section 3.

\subsection{The Optical Colours of BCGs}

Figure \ref{fig:plot_g_r_BCG_color_vs_cluster_mass_kcorr} shows that there is no evident correlation between the $(g-r)$ colour and the cluster halo mass, supporting the notion that the mass of the cluster in which the BCG resides does not affect its age. We repeat the same exercise studying the correlation between $(g-r)$ colour and stellar mass in our stellar mass complete sample and find weak correlations at all redshifts considered in this work. The Spearman rank correlation coefficient is $\rho = 0.4, 0.3, 0.1$ at $z = 0.1, 0.2, 0.3$, respectively, with probabilities of less than 1\% that the two quantities are uncorrelated. Interestingly, the BCGs in the blue tail in Fig. \ref{fig:plot_g_r_BCG_color_vs_cluster_mass_kcorr} do not have reliable stellar masses and thus do not enter the stellar mass complete sample: the weak correlation of the $(g-r)$ colour with stellar mass is therefore not driven by blue BCGs. 

Blue BCGs, defined here as galaxies $2\sigma$ bluer than the median $(g-r)$ colour, represent $6\%$ of our optically and $z_{phot}$ selected sample (not the sub-samples with stellar masses or WISE photometric data), with no significant trend with redshift, implying that their exclusion from the stellar mass complete sample does not bias our results.

Our result on the trend of the blue fraction with redshift appears in disagreement with \cite{Pipino_2011}, who showed that the fraction of blue BCGs increased with redshift at $z<0.35$ in a catalogue of clusters detected with the Adaptive Match Filter algorithm (\citealt{Szabo_2011}) in the SDSS DR6 (\citealt{Adelman_SDSS_DR6_2008}). We attribute this difference to the different cluster-search algorithms adopted for producing the catalogues used in the two papers, as well as to our conservative cuts in photometric redshift and photometry quality which reduced the number of galaxies at $z<0.35$ from 74,000 to 19,000. The cuts resulted in the loss of objects especially at high redshifts, where the quality of the photometry and the reliability of the photometric redshifts become lower.

The WHL15 catalogue was built using an algorithm that detected clusters in photometric redshift space. Although it was tested with visual inspection by the authors, it is still likely that it includes spurious systems which 
are 
not necessarily gravitationally bound. 
We matched our BCG sample with the {\ttfamily{redMaPPer}} and GMBCG cluster catalogues (\citealt{Rykoff_2014}, \citealt{Hao_2010}), which are cluster catalogues obtained with red sequence based cluster-search algorithms. In the 2,319 galaxies with rest-frame photometry in common with the {\ttfamily{redMaPPer}} catalogue, we find that the blue BCGs constitute 7\% of the population at $0.05 \leq z < 0.15$ and $3-4\%$ of the population at $0.15 < z < 0.35$. When considering the GMBCG catalogue, of the 3,123 BCGs in common with our catalogue, only $2-4\%$ are blue. 
The comparison with the {\ttfamily{redMaPPer}} and GMBCG cluster catalogues shows that the algorithms used to build those two catalogues can miss a significant fraction of the population of blue BCGs. 
As also noted in \cite{Pipino_2011}, the type of algorithm adopted for galaxy cluster detection is decisive in assessing the relevance of blue BCGs in the overall BCG population: 
algorithms based on colour selections that privilege red galaxies may miss blue BCGs and the galaxy clusters that host them.

In order to test what is the chance that random overdensities in photometric redshift are detected as clusters in our sample, we estimated the difference between photometric and spectroscopic redshifts $\Delta z = |z_{\rm phot} - z_{\rm spec}| / (1+z_{\rm spec})$ for the BCGs with spectroscopic redshifts in the three sub-samples of clusters studied in this paper. We find that the biweight median of this quantity never exceeds 0.1 at $0.05 \leq z < 0.35$. In less than 1\% of the cases $|z_{\rm phot} - z_{\rm spec}|$ is larger than 0.08. These results ensure that the chance that a cluster is a random overdensity of galaxies in photometric redshift space is less than 1\%. 
The results presented in this paper are therefore robust to cluster false detection.


\subsection{The Star Formation Activity in BCGs}

We discuss here the star-formation in our sample of BCGs, while the problem of the relationships between star-forming and blue BCGs is addressed in Section 4.4.

We use WISE to detect candidate star-forming and AGN-host BCGs and distinguish them from the quiescent ones. The advantage of using WISE with respect to ultra-violet colours as in \cite{Pipino_2011} is that the latter may be biased against dust-enshrouded star formation.

We could detect only 1,857 sources in the $W1$, $W2$ and $W3$ photometric bands and classify them as quiescent, star-forming or AGN hosts. These correspond to $\sim 10$\% of the sample that we study in the paper. The number of galaxies detected in the 3 filters at $z=0.25-0.35$ drops to $12$\% of the number of detections in the lower-redshift bins. This drop is driven by the shallow flux limit in the $W3$ band (In regions that are not confusion limited, AllWISE achieves $S/N=5$ flux at 54, 71, 730 and 5,000 mJy, corresponding to 16.9, 16.0, 11.5 and 8.0 mag) in $W1$, $W2$, $W3$ and $W4$, respectively\footnote{\url{http://wise2.ipac.caltech.edu/docs/release/allwise/expsup/sec2_1.html}}).

We detect star formation in $9$\% of the galaxies that have photometry in at least one of the WISE bands, implying that star formation in BCG is rare. With this result we extend the conclusions of \cite{Fraser_McKelvie_2014} on the rarity of star formation in BCGs to $z=0.35$ and with a sample of clusters that covers 1.5 orders of magnitude in halo mass. In Figure \ref{fig:plot_WISE_type_fractions} we also see that star formation in BCGs occurs more frequently at low halo masses in our sample and is more frequent at higher redshifts. The middle-left panel of Figure \ref{fig:plot_WISE_type_fractions} also shows that the fraction of star-forming BCGs strongly decreases with stellar mass. 

We note that in Tables 6-10 the estimated fractions can sum up to $>1$, especially in the lowest-populated bins. The approach that we use to derive the fractions and their uncertainties is based on the formalism presented in \cite{Dagostini_2004} and \cite{Cameron_2011} and it consists in applying the Bayes Theorem to a Binomial distribution with $n$ trials and $k$ successes. This approach allows one to estimate fractions of a given galaxy type even in the extreme cases in which $n=1$ and $k=0$. In this paper there are cases in which we can have at the same time no galaxies classified either as star-forming or AGN. In both cases $k=0$ and so we derive the same values for the fractions. These extreme cases, which lead to total fractions $>1$, should be interpreted as the {\itshape{probabilities}} of having star-forming or AGN-host galaxies in a given bin.

Our results on the trend of the fraction of star-forming BCGs with halo mass are in agreement with \cite{Oliva_Altamirano_2014}, who observed that $f_{\rm SF}$ is higher in low halo mass clusters in the GAMA survey. Furthermore, \cite{Gozaliasl_2016}, who studied BCGs and BGGs in a sample of clusters and groups with halo masses in the range $5\times 10^{12} - 10^{14.5} M_\odot$ at $z<1.0$, also found that the fraction of star-forming BCGs and BGGs decreases with the mass of the halo.

These authors also find that star formation increases with redshift, in agreement with \cite{Webb_2015}, who studied a sample of BCGs at $0.2 < z < 1.8$ from the Spitzer Adaptation of the Red Sequence Cluster Survey (SpARCS, \citealt{Wilson_2009}, \citealt{Muzzin_2009,Muzzin_2012}, \citealt{Demarco_2010_SpARCS}). \cite{Webb_2015} also find that below $z=0.6$ the BCGs classified as star-forming according to their IR flux ratios ($\rm 8.1 \mu m/4.5 \mu m$ vs $\rm 5.8 \mu m / 3.6 \mu m$) are rare.

\cite{Oliva_Altamirano_2014} suggest that the trends of $f_{\rm SF}$ with cluster mass may be a reflection of the correlation between BCG stellar mass and the mass the clusters. In order to test this, in Figure \ref{fig:plot_WISE_type_fractions} we split the sub-sample of BCGs which simultaneously have detections in W1, W2 and W3 and stellar masses in two classes of low- and high-stellar mass BCGs (bottom right and left panels in the figure). We set the boundary between high and low stellar masses at $M_*=10^{11.5} M_\odot$ and studied the fractions of quiescent and star-forming BCGs with halo mass. We see that the trends of $f_{\rm Q}$ and $f_{\rm SF}$ and $f_{\rm AGN}$ with $M_{200}$ are maintained for the lower-mass BCGs, while for higher-mass BCGs the trends of $f_{\rm Q}$, $f_{\rm SF}$ are flatter up to $M_{200} = 10^{14.2}M_\odot$ and then similar to the low stellar mass case at higher cluster masses. Thus the trend of the fractions of quiescent and star-forming galaxies with cluster mass are not strongly influenced by the $M_*$-$M_{200}$ correlation. This result is in agreement with Figure \ref{fig:plot_M200_vs_Mstar}, where it can be seen that our sample is characterised by a weaker correlation with respect to scaling relations published in the literature.

The middle-right panel of Figure \ref{fig:plot_WISE_type_fractions} shows that when one puts together the low and high stellar mass BCG, the trends of the star-forming and quiescent fractions with cluster mass is weakened, at least in the lowest halo mass bins. The trends resemble those observed in BCGs with $11.0 \leq \log(M_*/M_\odot) < 11.5$, reflecting the fact that the stellar mass complete sample is dominated by low-mass BCGs. 

The middle-left panel of Figure \ref{fig:plot_WISE_type_fractions} shows $f_{\rm Q}$ and $f_{\rm SF}$ plotted against BCG stellar mass. It can be seen that the dependence of the fractions of quiescent and star-forming galaxies with $M_*$ is stronger than that with cluster $M_{200}$. In order to test whether the trend is driven by the halo mass dependence of the fractions, we grouped clusters according to their halo masses and split the data-set in two sub-samples of clusters with masses lower and higher than the median of the halo mass range. As it can be seen in Table \ref{table10} the trend of the fraction of star-forming BCGs with the stellar mass of the galaxies remains even if we split in bins of halo mass. All these results support the notion that both stellar mass and cluster mass are important in regulating the star-formation activity in BCGs.

\cite{Runge_2018} selected 389 BCGs with $W4$ emission in the GMBCG catalogue at $0.1 < z < 0.55$. Most of these galaxies ($\sim 80$\%) are star-forming and only a minority of them are AGN hosts. We find only 83 galaxies in our sample that are in common with the \cite{Runge_2018} $W4$ emitters catalogue, mainly in low-mass clusters and groups. We attribute this low number of matches to the narrower range in redshift that we consider in this work ($z=0.05-0.35$ vs $z=0.1-0.55$) and to the fact that \cite{Runge_2018} performed their search for $W4$ emitters in the UnWISE images and the SDSS-WISE forced photometry catalogue (\citealt{Lang_2014_unWISE,Lang_2014_forced_photometry}), which deliver better resolution and signal-to-noise in $W3$ and $W4$ compared with the AllWISE catalogue.

\subsection{AGN Detection in BCGs at IR Wavelengths}

Figures \ref{fig:colour-colour_plots} and \ref{fig:plot_WISE_type_fractions} also show that there are overall few AGN host galaxies detected with WISE in our sample of BCGs. This result is in agreement with \cite{Webb_2015}. In our sample AGN hosts represent 1\% of the galaxies simultaneously detected in $W1$, $W2$ and $W3$, in agreement with \cite{Webb_2015} and \cite{Runge_2018}. We note that \cite{Oliva_Altamirano_2014} show that the AGN hosts represent $20-30\%$ in their sample of BCGs and that \cite{von_der_Linden_2007} show that the fraction of radio-loud AGN reaches up to 30\% at $z<0.1$. AGN in \cite{Oliva_Altamirano_2014} were selected optically, while in \cite{von_der_Linden_2007} they were selected at radio frequencies (1.4 GHz). \cite{Stern_2012} stress that the colour cut at $\rm (W1-W2) > 0.8$ used in this paper to identify AGN hosts can only select objects whose IR emission is dominated by the central AGN. Those authors show that if the contribution from the AGN to the IR continuum emission of the galaxy goes below 75\%, the $(W1-W2)$ colours become lower than 0.8. 

\begin{figure}
  \centering
	\includegraphics[width=0.5\textwidth, trim=0.0cm 0.0cm 0.0cm 0.0cm, clip, page=1]{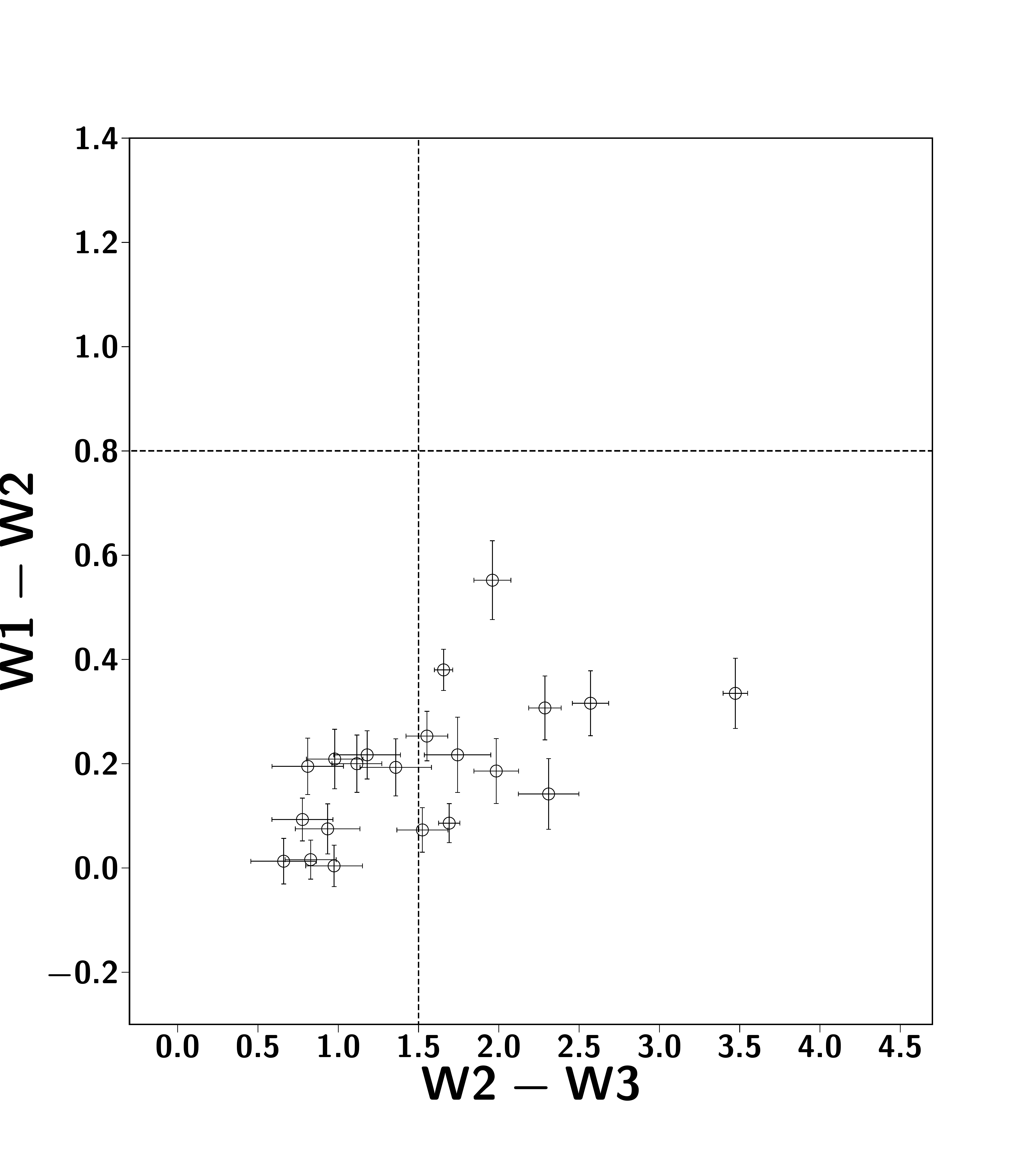}
	\caption{WISE 
	colour-colour diagram of the BCGs in common with the sample of radio-loud AGN from \protect\cite{Best_2005}. None of the BCGs that matches a radio-loud AGN in the \protect\cite{Best_2005} sample has $\rm (W1-W2) > 0.8$. Interestingly 11 of the 21 sources in common fall in the star-forming region of the colour-colour diagram.}
\label{fig:plot_BCG_WISE_color_color_plot_Best_et_al_2005_sources}
\end{figure}

A match of our sub-sample of WISE selected BCGs with the catalogue of radio-loud AGNs of \cite{Best_2005} detected in the SDSS DR2 shows that of the 19 galaxies common to both samples which have classification according to their WISE colours all have $\rm (W1-W2) < 0.8$. This result supports the notion that the WISE colour-colour classification adopted in this work is able to detect only the galaxies in which the IR emission is dominated by the central AGN. As it can be seen from Figure \ref{fig:plot_BCG_WISE_color_color_plot_Best_et_al_2005_sources}, 11 of the 19 BCGs in the radio-loud AGN sample of \cite{Best_2005} fall in the star-forming region of the colour-colour diagram. The study of AGN is beyond the scope of this work, and we only stress that selecting AGN with WISE in BCGs underestimates the fraction of active galaxies in the entire population. 


\subsection{Blue BCGs: are they all star forming?}

We find that $\sim$84\% of the BCGs that have detections in $W1$, $W2$ and $W3$ 
and that are $2\sigma$ bluer than the median $g-r$ colour are star-forming. This result implies that at least in the cases in which we have reliable IR photometry, star-forming BCGs make up the majority of the population in the blue tail of the $g-r$ vs $\rm M_{200}$ diagram. 
There still remains a part of these galaxies that are not classified as star-forming, suggesting that we need deeper data to better characterise the star formation activity of these objects (e.g.: unWISE).

We stress that only 10\% of the BCGs in our sample (see selection 
criteria above in Sect.2) are detected in $W1$, $W2$ and $W3$, and we cannot infer anything regarding star formation from the optical photometry alone and without reliable IR fluxes. Nevertheless, the colours of some of these galaxies are consistent with those in the blue cloud, suggesting that they may host star formation. 

\begin{figure*}
  \centering
	\includegraphics[width=\textwidth, trim=0.0cm 15.0cm 0.0cm 14.0cm, clip, page=1]{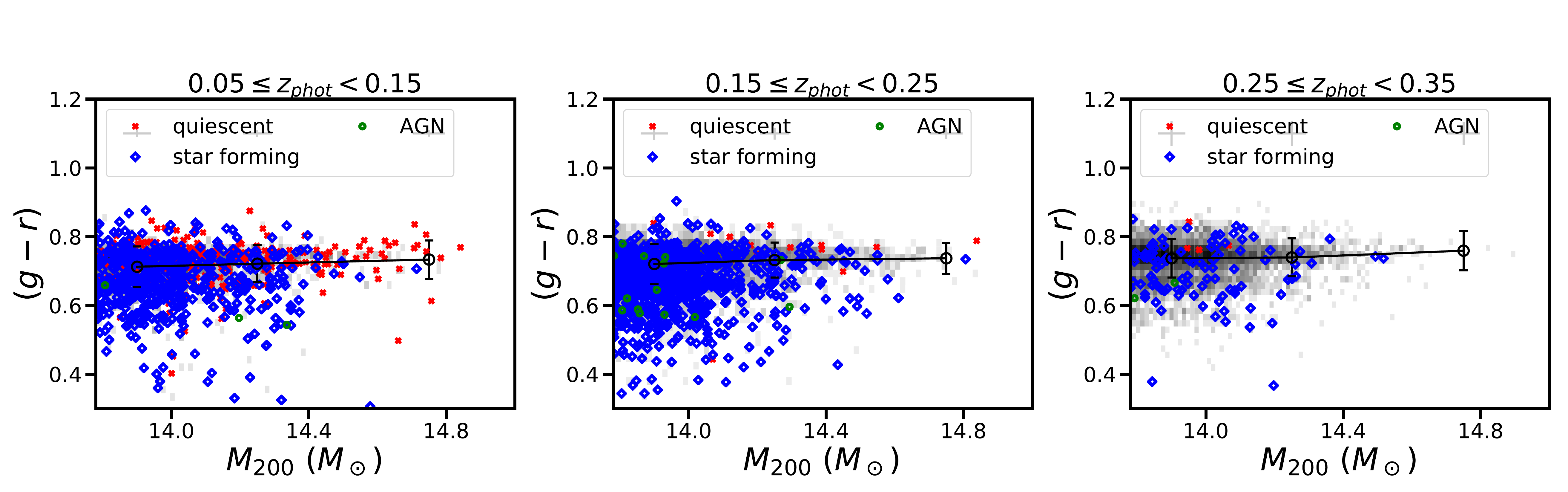}
	\caption{The same as Figure \ref{fig:plot_g_r_BCG_color_vs_cluster_mass_kcorr} but with galaxy types highlighted. While quiescent BCGs have optical colours consistent with red sequence galaxies, star-forming BCGs span a broad range of colours from blue to red. AGN represent a minority in our sample. The underlying grey area corresponds to the BCGs prior to the selection in $W1$, $W2$ and $W3$.}
\label{fig:plot_g_r_BCG_color_vs_cluster_mass_kcorr_with_WISE_type}
\end{figure*}

Figure \ref{fig:plot_g_r_BCG_color_vs_cluster_mass_kcorr_with_WISE_type} shows that while quiescent BCGs have all red $(g-r)$ colours, star-forming BCGs may have both blue and red colours. 
This suggests that there are BCGs where star formation is obscured by dust.

The origin of star formation in BCGs has been the subject of many works in the literature. For example, \cite{Rafferty_2008} showed that cooling flows are responsible for igniting star formation in BCGs, a conclusion that is in agreement with \cite{Pipino_2011}, but at odds with the more recent work of \cite{Runge_2018}.

In order to show the effects of cooling flows in our sample of BCGs, we matched our catalogue with the Archive of Chandra Cluster Entropy Profile Tables (ACCEPT, \citealt{Cavagnolo_2009}) and obtained a sub-sample of 60 sources detected within a radius of 10'', approximately corresponding to projected physical separations of 10 kpc and 50 kpc at $z=0.05$ and $z=0.35$, respectively. These numbers are consistent with the distances of the BCGs from their host cluster centres quoted in Table 1 of \cite{Bildfell_2008}.

\begin{figure}
  \centering
	\includegraphics[width=0.5\textwidth, trim=0.0cm 0.0cm 0.0cm 0.0cm, clip, page=1]{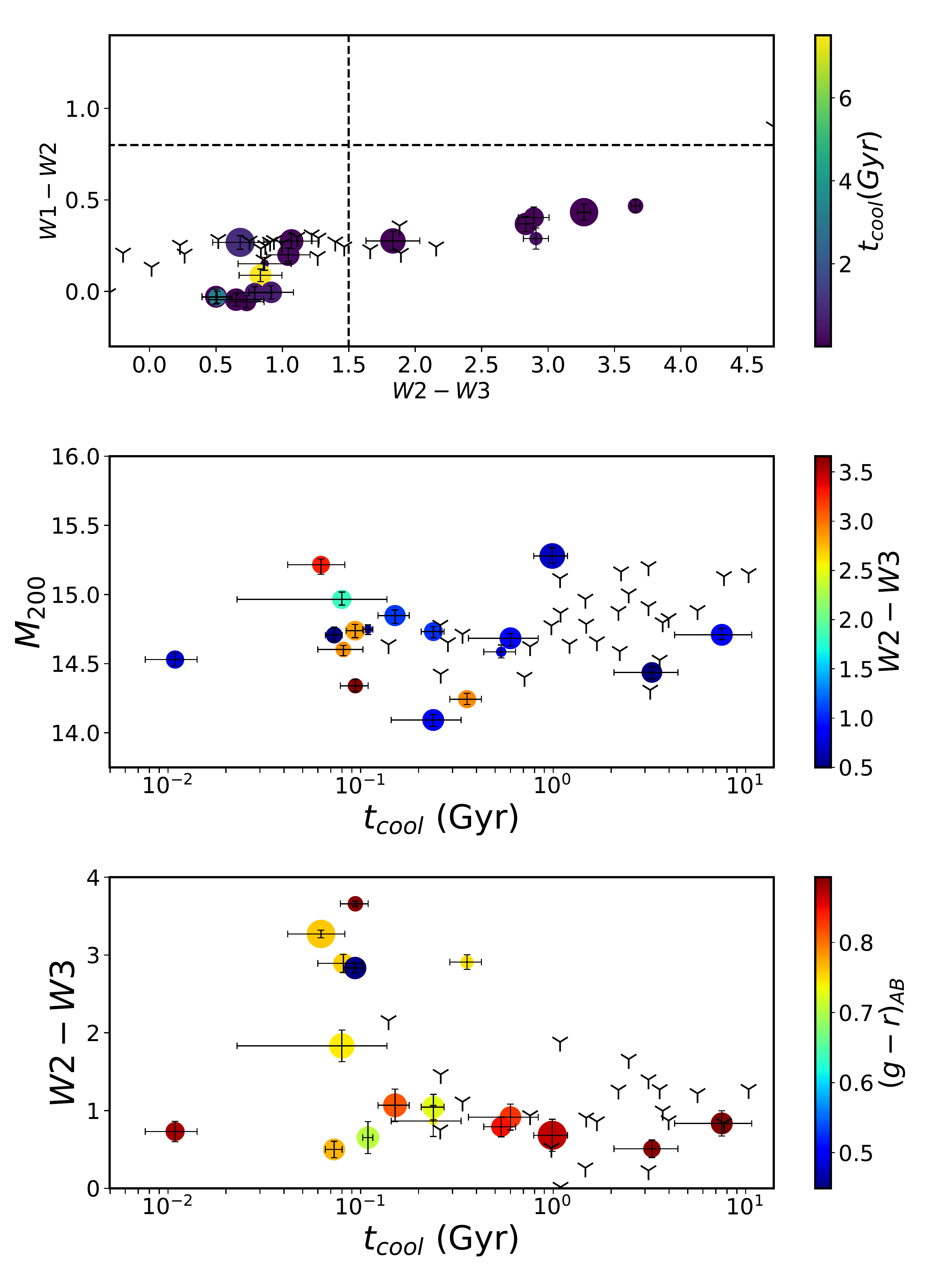}
	\caption{The effect of cooling flows on the star formation in BCGs. The plots in this figure are for a sub-sample of BCGs with sepctroscopic redshifts in the range $0.05 \leq z_{spec} < 0.35$ matched with the ACCEPT cluster catalogue. Black symbols represent upper limits in WISE photometry (see text). {\itshape{Top:}} WISE colour-colour diagram. Symbols are colour-coded according to the value of $t_{cool}$ and scaled proportinally to cluster halo mass. {\itshape{Middle}}: cluster halo mass as a function of $t_{cool}$. Symbols are colour-coded according to $W2-W3$ colours and scaled proprtionally to Petrosian $r$-band rest-frame magnitude. {\itshape{Bottom}}: $W2-W3$ colour as a function of $t_{cool}$. Symbols are colour-coded according to their rest-frame $g-r$ colours and scaled proportionally to cluster halo mass.}
\label{fig:plot_tcool}
\end{figure}

We found that only one object in the ACCEPT-matched catalogue passed the selection in SDSS and WISE photometry quality and photometric redshift quality (a quiescent galaxy at $z=0.14$). We therefore decided to relax our constraints on photometry and required that BCGs had spectroscopic redshifts and measured $W1-W2$ and $W2-W3$ colours to study their properties as a function of the cooling flow state of their host clusters. At $0.05 \leq z_{\rm spec} < 0.35$ there are 17 BCGs with $M_{200} \geq 6 \cdot 10^{13} M_\odot$ and measured IR colours. For the BCGs that are in the same range of $z_{\rm spec}$ and $M_{200}$ but do not have IR colours we derived upper limits for $W1-W2$ and $W2-W3$.

We estimated the cooling time $t_{cool}$ in years for each ACCEPT cluster using Equation 8 of \cite{Cavagnolo_2009}:
\begin{equation}
    t_{cool}(r) = t_{c0} + t_{100}\left(\frac{r}{100 \rm{ } kpc}\right)^\alpha 
\end{equation}

where $r$ is the projected distance (in kpc) from the cluster centroid, $t_{c0}$ is the cooling time in the core of the cluster, and $t_{100}$ is a normalisation (see \citealt{Cavagnolo_2009} for details). 

The top panel in Fig. \ref{fig:plot_tcool} shows the WISE colour-colour diagram with symbols colour-coded according to the values of $t_{cool}$ in Gyr and scaled as a function of $M_{200}$. It can be seen that galaxies with $(W2-W3) \geq 1.5$ tend to be found in clusters with $t_{cool} < 1 \, $Gyr, while BCGs with $(W2-W3) < 1.5$ tend to reside in clusters with a broader range of cooling times. The BCGs with upper limits on the IR photometry reside for their majority in the quiescent region of the diagram. We do not find any BCG with $(W1-W2) eq 0.8$ (AGN hosts) in the ACCEPT-matched sample.

The middle panel of Fig. \ref{fig:plot_tcool} presents $M_{200}$ plotted as a function of $t_{cool}$ with symbols colour coded  according to their $W2-W3$ colours and scaled as a function of $r$-band Petrosian magnitude. We used this quantity as a proxy for the BCG stellar mass as there are only 7 galaxies in the ACCEPT-matched sample that have measured stellar masses and IR colours at $0.05 \leq z_{spec} < 0.35$ and $M_{200} > 6 \cdot 10^{13} M_\odot$.

There is no apprent correlation between $M_{200}$ and $t_{cool}$. Clusters with $t_{cool} < 1.0$ Gyr host BCGs with a broad range of $W2-W3$ colours, while at higher cooling times we almost only find BCGs with $(W2-W3) < 1.5$. Almost all the BCGs with upper limits have $t_{cool} > 1.0$ Gyr, regardless of the masses of their host clusters. These results, together with the fact that the objects with upper limits reside in the quiescent region of the colour-colour diagram support the notion that the presence of star formation in BCGs is related with the cooling of the in-falling intra-cluster medium.

The bottom panel of Figure \ref{fig:plot_tcool}, where we plot the $W2-W3$ as a function of $t_{cool}$, explicitly shows this relationship. Clusters with $t_{cool} < 1$ Gyr have a broad range of $W2-W3$ values, while at $t_{cool} > 1$ Gyr the majority of BCGs, including those with only upper limits, have $W2-W3$ consistent with those of quiescent galaxies.

Interestingly, we notice in the middle panel of the figure that BCGs with lower $r-$band luminosities tend to lie in clusters with low cooling times, suggesting that these two quantities may be correlated.

The interpretation of the data shown in 
Fig. \ref{fig:plot_tcool} is coherent with the  hypothesis 
that recent star formation in BCGs originated as a consequence of cooling flows and that the occurrence of this environmentally-induced star formation does not depend on the halo mass of the clusters. 
These results are consistent with the findings of several other works ( e.e.:\ \citealt{Hu_1985}, \citealt{Bildfell_2008}, \citealt{Cavagnolo_2008}, \citealt{Pipino_2009}, \citealt{Donahue_2010, Donahue_2015}, \citealt{Loubser_2016}), which find that cool-core clusters are more likely to host BCGs with ongoing or recent star formation.

Star formation would therefore result as a consequence of the fact that the intra-cluster gas that falls in to the central galaxy is also cooling, allowing molecular 
hydrogen to form and fuel the formation of new stars. The cooling of gas affects only a minority of clusters in the local Universe and is regulated by a combination of pre-heating during cluster formation and AGN feedback from the centre of the BCG 
(see \citealt{McCarthy_2004, McCarthy_2008}).

Our results also suggest that cooling flows are a necessary condition for the onset of star formation in BCGs, i.e.: clusters with cooling flows may host BCGs with or without star formation; however, clusters without cooling flows will not host star-forming BCGs. We do not know why there are non star-froming BCGs in cool-core clusters nor is it possible to study that with the sample used in this paper. It may be related to mechanisms internal to the galaxies (e.g.: an AGN that heats or pushes the accreted gas out of the galaxy, see e.g:\ \citealt{McCarthy_2004, McCarthy_2008}) or these BCGs may have recently finished to form stars.

We refrain from any strong conclusion on this point which, with the data-set used would be speculative, and we just stress that our results on the IR colours suggest that star formation in BCGs is the result of processes that involve the stellar mass of the galaxies, the mass of the clusters and the cooling of the intra-cluster medium.

\subsection{Stellar Mass Growth}

Figure \ref{fig:plot_BCG_stellar_pop_Portsmouth_biweight_estimators_Wen_et_al_2015} presents the biweight locations and scale for the stellar masses of the BCGs as a function of redshift for each cluster subsample. We notice that at all redshifts the biweight locations of the stellar mass appear slightly shifted towards higher values when the mass of the clusters increases. This is an effect of the well known stellar mass vs halo mass correlation of BCGs (e.g.\ \citealt{Lidman_2012}, \citealt{Oliva_Altamirano_2014}, \citealt{Gozaliasl_2016}, \citealt{Bellstedt_2016}, \citealt{Lavoie_2016}). 

In Figure \ref{fig:plot_M200_vs_Mstar} we plot the cluster halo mass $M_{200}$ as a function of the stellar mass $M_*$ of the BCGs in the sub-sample of galaxies with reliable stellar masses at $0.05 \leq z < 0.35$. We also plot some of the results for the scaling relations between $M_{200}$ and $M_*$ found in the recent literature as well as the median and $1\sigma$ dispersion values of $M_{200}$ in bins of stellar mass. It can be seen that the median $M_{200}$ follows a significantly flatter trend with stellar mass with respect to the published scaling relations. We attribute this behaviour to the rigid limits that we set on cluster mass and BCG stellar mass to build complete samples in these two quantities. The study of the $M_{200}$ - $M_*$ scaling relation is beyond the scope of the present paper, and will be addressed in Orellana et al. (in prep.).

We note that in all the sub-samples stellar mass appears constant with redshift, indicating no growth with cosmic time at $z<0.35$. This result, which is in agreement with other observational works on low-redshift BCGs (e.g.\ \citealt{Lin_2013}, \citealt{Oliva_Altamirano_2014}, \citealt{Gozaliasl_2016}) appears in tension with some theoretical models that have been proposed for the formation and evolution of BCGs. For instance, in the model of \cite{De_Lucia_and_Blaizot_2007} BCGs should have accreted half of their mass since $z=0.4$, which we do not see in this paper. Also the model of \cite{Tonini_2012} shows an increase in the stellar mass of BCGs at low redshift, which we do not see in our sample. The models used in \cite{Contini_2018} for the stellar mass evolution of BCGs predict a trend that is closer to our constant trend at low redshifts. 

\cite{Oliva_Altamirano_2014} point out that the observed stellar masses from their and other literature works that they compare with in their paper are consistently below the predictions of the hierarchical models of \cite{De_Lucia_and_Blaizot_2007} and \cite{Tonini_2012}. Interestingly, the models of \cite{Contini_2018}, which take into account the intracluster light (ICL) in the treatment of the evolution of stellar mass in BCGs, predict an almost flat trend of the stellar mass with redshift at $z<0.3$. ICL is originated during the accretion of satellite galaxies on to the BCG (see \citealt{Rudick_2010}) and so the production of ICL could be responsible for the decline in the stellar mass growth of BCGs at low redshifts. We do not address here the study of the ICL, which would require an accurate and separate analysis and simply state that, in agreement with other works in the literature, our results suggest that there is no significant growth in the stellar mass of BCGs at $z<0.35$. We extend this conclusion to a large sample of clusters and in different halo mass regimes.

We remind the reader that we built the sub-samples of BCGs taking into account the accretion history of clusters (Section \ref{cluster_subsamples}). Our conclusions on stellar mass evolution are therefore robust to the effects of progenitor bias (\citealt{Lidman_2012}, \citealt{Oliva_Altamirano_2014}, \citealt{Shankar_2015}).

\section{Summary and Conclusions} \label{conclusions}

We presented a descriptive study of the optical and IR colours and stellar masses in a large and complete sample of BCGs
at $0.05 \leq z <0.35$ in the WHL15 galaxy clusters catalogue drawn from the Sloan Digital Sky Survey.

After selecting the objects with reliable photometry and photometric redshifts, we find that the optical rest-frame $g-r$ colours of the majority of the BCGs are consistent with those of quiescent red sequence galaxies. We detect a tail of blue BCGs at low cluster masses which represent the $6\%$ of the entire population. The $g-r$ colours of these galaxies are consistent with those of galaxies in the blue cloud and transition zone (i.e., the so-called green valley) of the colour-magnitude diagram.

The analysis of the WISE IR colours reveals that a small fraction (9\%) of the entire sample of BCGs are star-forming, supporting the notion that star-forming BCGs represent a minority in the entire BCG population. This result is in agreement with previous results on smaller samples at similar redshifts. We find that star-forming BCGs tend to be located in low halo mass clusters and groups and are more frequent at high redshifts. The fractions of star-forming and quiescent BCGs appear correlated slightly more strongly with stellar mass than with cluster halo mass. 

By matching 
our BCG sample
with the ACCEPT X-ray cluster catalog, we show that the presence of star formation in BCG is related with the existence of cooling flows in clusters, regardless of the halo mass of the clusters. Our results support a scenario in which cooling flows represent a necessary condition for the onset of star formation in BCGs: cool-core clusters may host both star-forming and quiescent BCGs; however, non cool-core clusters only host quiescent BCGs. We find hints that the $r-$band luminosity of the BCGs and the cooling time of the intra-cluster medium may be correlated.

Overall, star formation in BCGs appears to be the result of physical processes that depend on the stellar mass of the galaxies, the halo mass of the clusters and the presence of cooling flows.

The IR colours also reveal that 84\% of the optically blue BCGs detected in $W1$, $W2$ and $W3$ are star-forming, suggesting that in the majority of the cases the blue colours are the result of ongoing star formation. We stress that this conclusion only takes into account the galaxies that have reliable photometry in the $W1$, $W2$ and $W3$ WISE IR bands, which represent 10\% of our sample. Better quality data such as unWISE and the SDSS spectroscopic database need to be explored in order to get a better census of star formation in BCGs. 

Interestingly, while quiescent BCGs have mostly red optical colours, star-forming BCGs may have both blue and red colours, suggesting that star formation may be obscured by dust in some of them. We find a small number of AGN detected with the WISE colour-colour classification ($\sim 1\%$ of the sub-sample with photometry in $W1$, $W2$ and $W3$).

The optical colours of BCGs do not appear correlated with cluster mass, although they are weakly correlated with the stellar mass of the galaxies, suggesting that stellar mass is more important than cluster mass in determining the ages of the stellar populations in BCGs.

The analysis of the stellar masses suggests that there is no significant growth in the redshift range considered in this work, in agreement with other observational results in the literature.

The comparisons with the RedMapper and GMBCG samples show that these cluster-search algorithms can miss an important fraction of blue and star-forming BCGs. Studies of star-forming BCGs relying on these samples may be affected by significant biases.

Forthcoming papers in this series will focus on particular aspects of BCGs (star formation, AGN, stellar populations) and investigate them in the context of the clusters in which reside.

\section*{Acknowledgements}

We thank the anonymous referee for the helpful and constructive feedback. P.C. acknowledges the support provided by FONDECYT postdoctoral research grant no 3160375. 
G.O. acknowledges the support provided by FONDECYT postdoctoral research grant no 3170942.
This publication makes use of data products from the Wide-field Infrared Survey Explorer, which is a joint project of the University of California, Los Angeles, and the Jet Propulsion Laboratory/California Institute of Technology, funded by the National Aeronautics and Space Administration. Funding for SDSS-III has been provided by the Alfred P. Sloan Foundation, the Participating Institutions, the National Science Foundation, and the U.S. Department of Energy Office of Science. The SDSS-III web site is http://www.sdss3.org/. SDSS-III is managed by the Astrophysical Research Consortium for the Participating Institutions of the SDSS-III Collaboration including the University of Arizona, the Brazilian Participation Group, Brookhaven National Laboratory, Carnegie Mellon University, University of Florida, the French Participation Group, the German Participation Group, Harvard University, the Instituto de Astrofisica de Canarias, the Michigan State/Notre Dame/JINA Participation Group, Johns Hopkins University, Lawrence Berkeley National Laboratory, Max Planck Institute for Astrophysics, Max Planck Institute for Extraterrestrial Physics, New Mexico State University, New York University, Ohio State University, Pennsylvania State University, University of Portsmouth, Princeton University, the Spanish Participation Group, University of Tokyo, University of Utah, Vanderbilt University, University of Virginia, University of Washington, and Yale University.





\bibliographystyle{mnras}
\bibliography{pcerulo_BCG_SDSS} 

\begin{thebibliography}{}
\makeatletter
\relax
\def\mn@urlcharsother{\let\do\@makeother \do\$\do\&\do\#\do\^\do\_\do\%\do\~}
\def\mn@doi{\begingroup\mn@urlcharsother \@ifnextchar [ {\mn@doi@}
  {\mn@doi@[]}}
\def\mn@doi@[#1]#2{\def\@tempa{#1}\ifx\@tempa\@empty \href
  {http://dx.doi.org/#2} {doi:#2}\else \href {http://dx.doi.org/#2} {#1}\fi
  \endgroup}
\def\mn@eprint#1#2{\mn@eprint@#1:#2::\@nil}
\def\mn@eprint@arXiv#1{\href {http://arxiv.org/abs/#1} {{\tt arXiv:#1}}}
\def\mn@eprint@dblp#1{\href {http://dblp.uni-trier.de/rec/bibtex/#1.xml}
  {dblp:#1}}
\def\mn@eprint@#1:#2:#3:#4\@nil{\def\@tempa {#1}\def\@tempb {#2}\def\@tempc
  {#3}\ifx \@tempc \@empty \let \@tempc \@tempb \let \@tempb \@tempa \fi \ifx
  \@tempb \@empty \def\@tempb {arXiv}\fi \@ifundefined
  {mn@eprint@\@tempb}{\@tempb:\@tempc}{\expandafter \expandafter \csname
  mn@eprint@\@tempb\endcsname \expandafter{\@tempc}}}

\bibitem[\protect\citeauthoryear{{Adelman-McCarthy} et~al.,}{{Adelman-McCarthy}
  et~al.}{2008}]{Adelman_SDSS_DR6_2008}
{Adelman-McCarthy} J.~K.,  et~al., 2008, \mn@doi [\apjs] {10.1086/524984},
  \href {http://cdsads.u-strasbg.fr/abs/2008ApJS..175..297A} {175, 297}

\bibitem[\protect\citeauthoryear{{Ahn} et~al.,}{{Ahn}
  et~al.}{2014}]{Ahn_SDSS_DR10_2014}
{Ahn} C.~P.,  et~al., 2014, \mn@doi [\apjs] {10.1088/0067-0049/211/2/17}, \href
  {http://cdsads.u-strasbg.fr/abs/2014ApJS..211...17A} {211, 17}

\bibitem[\protect\citeauthoryear{{Aihara} et~al.,}{{Aihara}
  et~al.}{2011}]{Aihara_2011}
{Aihara} H.,  et~al., 2011, \mn@doi [\apjs] {10.1088/0067-0049/193/2/29}, \href
  {http://cdsads.u-strasbg.fr/abs/2011ApJS..193...29A} {193, 29}

\bibitem[\protect\citeauthoryear{{Alam} et~al.,}{{Alam}
  et~al.}{2015}]{Alam_2015}
{Alam} S.,  et~al., 2015, \mn@doi [\apjs] {10.1088/0067-0049/219/1/12}, \href
  {http://cdsads.u-strasbg.fr/abs/2015ApJS..219...12A} {219, 12}

\bibitem[\protect\citeauthoryear{{Aragon-Salamanca}, {Baugh}  \&
  {Kauffmann}}{{Aragon-Salamanca} et~al.}{1998}]{Aragon_Salamanca_1998}
{Aragon-Salamanca} A.,  {Baugh} C.~M.,   {Kauffmann} G.,  1998, \mn@doi
  [\mnras] {10.1046/j.1365-8711.1998.01495.x}, \href
  {http://cdsads.u-strasbg.fr/abs/1998MNRAS.297..427A} {297, 427}

\bibitem[\protect\citeauthoryear{{Ascaso}, {Lemaux}, {Lubin}, {Gal},
  {Kocevski}, {Rumbaugh}  \& {Squires}}{{Ascaso} et~al.}{2014}]{Ascaso_2014}
{Ascaso} B.,  {Lemaux} B.~C.,  {Lubin} L.~M.,  {Gal} R.~R.,  {Kocevski} D.~D.,
  {Rumbaugh} N.,   {Squires} G.,  2014, \mn@doi [\mnras]
  {10.1093/mnras/stu877}, \href
  {http://cdsads.u-strasbg.fr/abs/2014MNRAS.442..589A} {442, 589}

\bibitem[\protect\citeauthoryear{{Baldry}, {Balogh}, {Bower}, {Glazebrook},
  {Nichol}, {Bamford}  \& {Budavari}}{{Baldry} et~al.}{2006}]{Baldry_2006}
{Baldry} I.~K.,  {Balogh} M.~L.,  {Bower} R.~G.,  {Glazebrook} K.,  {Nichol}
  R.~C.,  {Bamford} S.~P.,   {Budavari} T.,  2006, \mn@doi [\mnras]
  {10.1111/j.1365-2966.2006.11081.x}, \href
  {http://cdsads.u-strasbg.fr/abs/2006MNRAS.373..469B} {373, 469}

\bibitem[\protect\citeauthoryear{{Beck}, {Dobos}, {Budav{\'a}ri}, {Szalay}  \&
  {Csabai}}{{Beck} et~al.}{2016}]{Beck_2016}
{Beck} R.,  {Dobos} L.,  {Budav{\'a}ri} T.,  {Szalay} A.~S.,   {Csabai} I.,
  2016, \mn@doi [\mnras] {10.1093/mnras/stw1009}, \href
  {http://cdsads.u-strasbg.fr/abs/2016MNRAS.460.1371B} {460, 1371}

\bibitem[\protect\citeauthoryear{{Beers}, {Flynn}  \& {Gebhardt}}{{Beers}
  et~al.}{1990}]{Beers_1990}
{Beers} T.~C.,  {Flynn} K.,   {Gebhardt} K.,  1990, \mn@doi [\aj]
  {10.1086/115487}, \href {http://cdsads.u-strasbg.fr/abs/1990AJ....100...32B}
  {100, 32}

\bibitem[\protect\citeauthoryear{{Bellstedt} et~al.,}{{Bellstedt}
  et~al.}{2016}]{Bellstedt_2016}
{Bellstedt} S.,  et~al., 2016, \mn@doi [\mnras] {10.1093/mnras/stw1184}, \href
  {http://cdsads.u-strasbg.fr/abs/2016MNRAS.460.2862B} {460, 2862}

\bibitem[\protect\citeauthoryear{{Bernardi}, {Hyde}, {Sheth}, {Miller}  \&
  {Nichol}}{{Bernardi} et~al.}{2007}]{Bernardi_2007}
{Bernardi} M.,  {Hyde} J.~B.,  {Sheth} R.~K.,  {Miller} C.~J.,   {Nichol}
  R.~C.,  2007, \mn@doi [\aj] {10.1086/511783}, \href
  {http://cdsads.u-strasbg.fr/abs/2007AJ....133.1741B} {133, 1741}

\bibitem[\protect\citeauthoryear{{Best}, {Kauffmann}, {Heckman}  \&
  {Ivezi{\'c}}}{{Best} et~al.}{2005}]{Best_2005}
{Best} P.~N.,  {Kauffmann} G.,  {Heckman} T.~M.,   {Ivezi{\'c}} {\v Z}.,  2005,
  \mn@doi [\mnras] {10.1111/j.1365-2966.2005.09283.x}, \href
  {http://cdsads.u-strasbg.fr/abs/2005MNRAS.362....9B} {362, 9}

\bibitem[\protect\citeauthoryear{{Best}, {von der Linden}, {Kauffmann},
  {Heckman}  \& {Kaiser}}{{Best} et~al.}{2007}]{Best_2007}
{Best} P.~N.,  {von der Linden} A.,  {Kauffmann} G.,  {Heckman} T.~M.,
  {Kaiser} C.~R.,  2007, \mn@doi [\mnras] {10.1111/j.1365-2966.2007.11937.x},
  \href {http://cdsads.u-strasbg.fr/abs/2007MNRAS.379..894B} {379, 894}

\bibitem[\protect\citeauthoryear{{Bildfell}, {Hoekstra}, {Babul}  \&
  {Mahdavi}}{{Bildfell} et~al.}{2008}]{Bildfell_2008}
{Bildfell} C.,  {Hoekstra} H.,  {Babul} A.,   {Mahdavi} A.,  2008, \mn@doi
  [\mnras] {10.1111/j.1365-2966.2008.13699.x}, \href
  {http://cdsads.u-strasbg.fr/abs/2008MNRAS.389.1637B} {389, 1637}

\bibitem[\protect\citeauthoryear{{Bolzonella}, {Miralles}  \&
  {Pell{\'o}}}{{Bolzonella} et~al.}{2000}]{Bolzonella_2000}
{Bolzonella} M.,  {Miralles} J.-M.,   {Pell{\'o}} R.,  2000, \aap, \href
  {http://cdsads.u-strasbg.fr/abs/2000A%26A...363..476B} {363, 476}

\bibitem[\protect\citeauthoryear{{Bonaventura} et~al.,}{{Bonaventura}
  et~al.}{2017}]{Bonaventura_2017}
{Bonaventura} N.~R.,  et~al., 2017, \mn@doi [\mnras] {10.1093/mnras/stx722},
  \href {http://cdsads.u-strasbg.fr/abs/2017MNRAS.469.1259B} {469, 1259}

\bibitem[\protect\citeauthoryear{{Bower}, {Benson}, {Malbon}, {Helly}, {Frenk},
  {Baugh}, {Cole}  \& {Lacey}}{{Bower} et~al.}{2006}]{Bower_2006}
{Bower} R.~G.,  {Benson} A.~J.,  {Malbon} R.,  {Helly} J.~C.,  {Frenk} C.~S.,
  {Baugh} C.~M.,  {Cole} S.,   {Lacey} C.~G.,  2006, \mn@doi [\mnras]
  {10.1111/j.1365-2966.2006.10519.x}, \href
  {http://cdsads.u-strasbg.fr/abs/2006MNRAS.370..645B} {370, 645}

\bibitem[\protect\citeauthoryear{{Brough}, {Proctor}, {Forbes}, {Couch},
  {Collins}, {Burke}  \& {Mann}}{{Brough} et~al.}{2007}]{Brough_2007}
{Brough} S.,  {Proctor} R.,  {Forbes} D.~A.,  {Couch} W.~J.,  {Collins} C.~A.,
  {Burke} D.~J.,   {Mann} R.~G.,  2007, \mn@doi [\mnras]
  {10.1111/j.1365-2966.2007.11900.x}, \href
  {http://cdsads.u-strasbg.fr/abs/2007MNRAS.378.1507B} {378, 1507}

\bibitem[\protect\citeauthoryear{{Brough}, {Couch}, {Collins}, {Jarrett},
  {Burke}  \& {Mann}}{{Brough} et~al.}{2008}]{Brough_2008}
{Brough} S.,  {Couch} W.~J.,  {Collins} C.~A.,  {Jarrett} T.,  {Burke} D.~J.,
  {Mann} R.~G.,  2008, \mn@doi [\mnras] {10.1111/j.1745-3933.2008.00442.x},
  \href {http://cdsads.u-strasbg.fr/abs/2008MNRAS.385L.103B} {385, L103}

\bibitem[\protect\citeauthoryear{{Brown}, {Jarrett}  \& {Cluver}}{{Brown}
  et~al.}{2014}]{Brown_2014}
{Brown} M.~J.~I.,  {Jarrett} T.~H.,   {Cluver} M.~E.,  2014, \mn@doi [\pasa]
  {10.1017/pasa.2014.44}, \href
  {http://cdsads.u-strasbg.fr/abs/2014PASA...31...49B} {31, HASH}

\bibitem[\protect\citeauthoryear{{Bruzual} \& {Charlot}}{{Bruzual} \&
  {Charlot}}{2003}]{Bruzual_2003}
{Bruzual} G.,  {Charlot} S.,  2003, \mn@doi [\mnras]
  {10.1046/j.1365-8711.2003.06897.x}, \href
  {http://adsabs.harvard.edu/abs/2003MNRAS.344.1000B} {344, 1000}

\bibitem[\protect\citeauthoryear{{Cameron}}{{Cameron}}{2011}]{Cameron_2011}
{Cameron} E.,  2011, \mn@doi [\pasa] {10.1071/AS10046}, \href
  {http://adsabs.harvard.edu/abs/2011PASA...28..128C} {28, 128}

\bibitem[\protect\citeauthoryear{{Cavagnolo}, {Donahue}, {Voit}  \&
  {Sun}}{{Cavagnolo} et~al.}{2008}]{Cavagnolo_2008}
{Cavagnolo} K.~W.,  {Donahue} M.,  {Voit} G.~M.,   {Sun} M.,  2008, \mn@doi
  [\apjl] {10.1086/591665}, \href
  {http://cdsads.u-strasbg.fr/abs/2008ApJ...683L.107C} {683, L107}

\bibitem[\protect\citeauthoryear{{Cavagnolo}, {Donahue}, {Voit}  \&
  {Sun}}{{Cavagnolo} et~al.}{2009}]{Cavagnolo_2009}
{Cavagnolo} K.~W.,  {Donahue} M.,  {Voit} G.~M.,   {Sun} M.,  2009, \mn@doi
  [\apjs] {10.1088/0067-0049/182/1/12}, \href
  {http://cdsads.u-strasbg.fr/abs/2009ApJS..182...12C} {182, 12}

\bibitem[\protect\citeauthoryear{{Chang}, {van der Wel}, {da Cunha}  \&
  {Rix}}{{Chang} et~al.}{2015}]{Chang_2015}
{Chang} Y.-Y.,  {van der Wel} A.,  {da Cunha} E.,   {Rix} H.-W.,  2015, \mn@doi
  [\apjs] {10.1088/0067-0049/219/1/8}, \href
  {http://adsabs.harvard.edu/abs/2015ApJS..219....8C} {219, 8}

\bibitem[\protect\citeauthoryear{{Cluver} et~al.,}{{Cluver}
  et~al.}{2014}]{Cluver_2014}
{Cluver} M.~E.,  et~al., 2014, \mn@doi [\apj] {10.1088/0004-637X/782/2/90},
  \href {http://cdsads.u-strasbg.fr/abs/2014ApJ...782...90C} {782, 90}

\bibitem[\protect\citeauthoryear{{Contini}, {Yi}  \& {Kang}}{{Contini}
  et~al.}{2018}]{Contini_2018}
{Contini} E.,  {Yi} S.~K.,   {Kang} X.,  2018, \mn@doi [\mnras]
  {10.1093/mnras/sty1518}, \href
  {http://cdsads.u-strasbg.fr/abs/2018MNRAS.tmp.1448C} {}

\bibitem[\protect\citeauthoryear{{Contini}, {Yi}  \& {Kang}}{{Contini}
  et~al.}{2019}]{Contini_2019}
{Contini} E.,  {Yi} S.~K.,   {Kang} X.,  2019, \mn@doi [\apj]
  {10.3847/1538-4357/aaf41f}, \href
  {http://cdsads.u-strasbg.fr/abs/2019ApJ...871...24C} {871, 24}

\bibitem[\protect\citeauthoryear{{Correa}, {Wyithe}, {Schaye}  \&
  {Duffy}}{{Correa} et~al.}{2015a}]{Correa_2015a}
{Correa} C.~A.,  {Wyithe} J.~S.~B.,  {Schaye} J.,   {Duffy} A.~R.,  2015a,
  \mn@doi [\mnras] {10.1093/mnras/stv689}, \href
  {http://cdsads.u-strasbg.fr/abs/2015MNRAS.450.1514C} {450, 1514}

\bibitem[\protect\citeauthoryear{{Correa}, {Wyithe}, {Schaye}  \&
  {Duffy}}{{Correa} et~al.}{2015b}]{Correa_2015b}
{Correa} C.~A.,  {Wyithe} J.~S.~B.,  {Schaye} J.,   {Duffy} A.~R.,  2015b,
  \mn@doi [\mnras] {10.1093/mnras/stv697}, \href
  {http://cdsads.u-strasbg.fr/abs/2015MNRAS.450.1521C} {450, 1521}

\bibitem[\protect\citeauthoryear{{Correa}, {Wyithe}, {Schaye}  \&
  {Duffy}}{{Correa} et~al.}{2015c}]{Correa_2015c}
{Correa} C.~A.,  {Wyithe} J.~S.~B.,  {Schaye} J.,   {Duffy} A.~R.,  2015c,
  \mn@doi [\mnras] {10.1093/mnras/stv1363}, \href
  {http://cdsads.u-strasbg.fr/abs/2015MNRAS.452.1217C} {452, 1217}

\bibitem[\protect\citeauthoryear{{Covone}, {Sereno}, {Kilbinger}  \&
  {Cardone}}{{Covone} et~al.}{2014}]{Covone_2014}
{Covone} G.,  {Sereno} M.,  {Kilbinger} M.,   {Cardone} V.~F.,  2014, \mn@doi
  [\apjl] {10.1088/2041-8205/784/2/L25}, \href
  {http://adsabs.harvard.edu/abs/2014ApJ...784L..25C} {784, L25}

\bibitem[\protect\citeauthoryear{{Croft}, {de Vries}  \& {Becker}}{{Croft}
  et~al.}{2007}]{Croft_2007}
{Croft} S.,  {de Vries} W.,   {Becker} R.~H.,  2007, \mn@doi [\apjl]
  {10.1086/522086}, \href {http://cdsads.u-strasbg.fr/abs/2007ApJ...667L..13C}
  {667, L13}

\bibitem[\protect\citeauthoryear{{Croton} et~al.,}{{Croton}
  et~al.}{2006}]{Croton_2006}
{Croton} D.~J.,  et~al., 2006, \mn@doi [\mnras]
  {10.1111/j.1365-2966.2005.09675.x}, \href
  {http://cdsads.u-strasbg.fr/abs/2006MNRAS.365...11C} {365, 11}

\bibitem[\protect\citeauthoryear{{D'Agostini}}{{D'Agostini}}{2004}]{Dagostini_2004}
{D'Agostini} G.,  2004, ArXiv Physics e-prints, \href
  {http://cdsads.u-strasbg.fr/abs/2004physics..12069D} {}

\bibitem[\protect\citeauthoryear{{D'Souza}, {Vegetti}  \&
  {Kauffmann}}{{D'Souza} et~al.}{2015}]{DSouza_2015}
{D'Souza} R.,  {Vegetti} S.,   {Kauffmann} G.,  2015, \mn@doi [\mnras]
  {10.1093/mnras/stv2234}, \href
  {http://cdsads.u-strasbg.fr/abs/2015MNRAS.454.4027D} {454, 4027}

\bibitem[\protect\citeauthoryear{{De Lucia} et~al.,}{{De Lucia}
  et~al.}{2007}]{De_Lucia_and_Blaizot_2007}
{De Lucia} G.,  et~al., 2007, \mn@doi [\mnras]
  {10.1111/j.1365-2966.2006.11199.x}, \href
  {http://cdsads.u-strasbg.fr/abs/2007MNRAS.374..809D} {374, 809}

\bibitem[\protect\citeauthoryear{{Demarco} et~al.,}{{Demarco}
  et~al.}{2010}]{Demarco_2010_SpARCS}
{Demarco} R.,  et~al., 2010, \mn@doi [\apj] {10.1088/0004-637X/711/2/1185},
  \href {http://cdsads.u-strasbg.fr/abs/2010ApJ...711.1185D} {711, 1185}

\bibitem[\protect\citeauthoryear{{Donahue} et~al.,}{{Donahue}
  et~al.}{2010}]{Donahue_2010}
{Donahue} M.,  et~al., 2010, \mn@doi [\apj] {10.1088/0004-637X/715/2/881},
  \href {http://cdsads.u-strasbg.fr/abs/2010ApJ...715..881D} {715, 881}

\bibitem[\protect\citeauthoryear{{Donahue} et~al.,}{{Donahue}
  et~al.}{2015}]{Donahue_2015}
{Donahue} M.,  et~al., 2015, \mn@doi [\apj] {10.1088/0004-637X/805/2/177},
  \href {http://cdsads.u-strasbg.fr/abs/2015ApJ...805..177D} {805, 177}

\bibitem[\protect\citeauthoryear{{Driver} et~al.,}{{Driver}
  et~al.}{2011}]{Driver_2011}
{Driver} S.~P.,  et~al., 2011, \mn@doi [\mnras]
  {10.1111/j.1365-2966.2010.18188.x}, \href
  {http://cdsads.u-strasbg.fr/abs/2011MNRAS.413..971D} {413, 971}

\bibitem[\protect\citeauthoryear{{Dubinski}}{{Dubinski}}{1998}]{Dubinski_1998}
{Dubinski} J.,  1998, \mn@doi [\apj] {10.1086/305901}, \href
  {http://cdsads.u-strasbg.fr/abs/1998ApJ...502..141D} {502, 141}

\bibitem[\protect\citeauthoryear{{Eminian}, {Kauffmann}, {Charlot}, {Wild},
  {Bruzual}, {Rettura}  \& {Loveday}}{{Eminian} et~al.}{2008}]{Eminian_2008}
{Eminian} C.,  {Kauffmann} G.,  {Charlot} S.,  {Wild} V.,  {Bruzual} G.,
  {Rettura} A.,   {Loveday} J.,  2008, \mn@doi [\mnras]
  {10.1111/j.1365-2966.2007.12742.x}, \href
  {http://cdsads.u-strasbg.fr/abs/2008MNRAS.384..930E} {384, 930}

\bibitem[\protect\citeauthoryear{{Fasano} et~al.,}{{Fasano}
  et~al.}{2006}]{Fasano_2006}
{Fasano} G.,  et~al., 2006, \mn@doi [\aap] {10.1051/0004-6361:20053816}, \href
  {http://adsabs.harvard.edu/abs/2006A%26A...445..805F} {445, 805}

\bibitem[\protect\citeauthoryear{{Fraser-McKelvie}, {Brown}  \&
  {Pimbblet}}{{Fraser-McKelvie} et~al.}{2014}]{Fraser_McKelvie_2014}
{Fraser-McKelvie} A.,  {Brown} M.~J.~I.,   {Pimbblet} K.~A.,  2014, \mn@doi
  [\mnras] {10.1093/mnrasl/slu117}, \href
  {http://cdsads.u-strasbg.fr/abs/2014MNRAS.444L..63F} {444, L63}

\bibitem[\protect\citeauthoryear{{Gabor} \& {Dav{\'e}}}{{Gabor} \&
  {Dav{\'e}}}{2012}]{Gabor_2012}
{Gabor} J.~M.,  {Dav{\'e}} R.,  2012, \mn@doi [\mnras]
  {10.1111/j.1365-2966.2012.21640.x}, \href
  {http://cdsads.u-strasbg.fr/abs/2012MNRAS.427.1816G} {427, 1816}

\bibitem[\protect\citeauthoryear{{Gozaliasl}, {Finoguenov}, {Khosroshahi},
  {Mirkazemi}, {Erfanianfar}  \& {Tanaka}}{{Gozaliasl}
  et~al.}{2016}]{Gozaliasl_2016}
{Gozaliasl} G.,  {Finoguenov} A.,  {Khosroshahi} H.~G.,  {Mirkazemi} M.,
  {Erfanianfar} G.,   {Tanaka} M.,  2016, \mn@doi [\mnras]
  {10.1093/mnras/stw448}, \href
  {http://cdsads.u-strasbg.fr/abs/2016MNRAS.458.2762G} {458, 2762}

\bibitem[\protect\citeauthoryear{{Graham}, {Lauer}, {Colless}  \&
  {Postman}}{{Graham} et~al.}{1996}]{Graham_1996}
{Graham} A.,  {Lauer} T.~R.,  {Colless} M.,   {Postman} M.,  1996, \mn@doi
  [\apj] {10.1086/177440}, \href
  {http://adsabs.harvard.edu/abs/1996ApJ...465..534G} {465, 534}

\bibitem[\protect\citeauthoryear{{Hao} et~al.,}{{Hao} et~al.}{2010}]{Hao_2010}
{Hao} J.,  et~al., 2010, \mn@doi [\apjs] {10.1088/0067-0049/191/2/254}, \href
  {http://cdsads.u-strasbg.fr/abs/2010ApJS..191..254H} {191, 254}

\bibitem[\protect\citeauthoryear{{Hinshaw} et~al.,}{{Hinshaw}
  et~al.}{2009}]{Hinshaw_2009}
{Hinshaw} G.,  et~al., 2009, \mn@doi [\apjs] {10.1088/0067-0049/180/2/225},
  \href {http://cdsads.u-strasbg.fr/abs/2009ApJS..180..225H} {180, 225}

\bibitem[\protect\citeauthoryear{{Hogg} et~al.,}{{Hogg}
  et~al.}{2004}]{Hogg_2004}
{Hogg} D.~W.,  et~al., 2004, \mn@doi [\apjl] {10.1086/381749}, \href
  {http://adsabs.harvard.edu/abs/2004ApJ...601L..29H} {601, L29}

\bibitem[\protect\citeauthoryear{{Hu}, {Cowie}  \& {Wang}}{{Hu}
  et~al.}{1985}]{Hu_1985}
{Hu} E.~M.,  {Cowie} L.~L.,   {Wang} Z.,  1985, \mn@doi [\apjs]
  {10.1086/191081}, \href {http://cdsads.u-strasbg.fr/abs/1985ApJS...59..447H}
  {59, 447}

\bibitem[\protect\citeauthoryear{{Huchra} \& {Geller}}{{Huchra} \&
  {Geller}}{1982}]{Huchra_Geller_1982}
{Huchra} J.~P.,  {Geller} M.~J.,  1982, \mn@doi [\apj] {10.1086/160000}, \href
  {http://cdsads.u-strasbg.fr/abs/1982ApJ...257..423H} {257, 423}

\bibitem[\protect\citeauthoryear{{Jarrett} et~al.,}{{Jarrett}
  et~al.}{2011}]{Jarrett_2011}
{Jarrett} T.~H.,  et~al., 2011, \mn@doi [\apj] {10.1088/0004-637X/735/2/112},
  \href {http://adsabs.harvard.edu/abs/2011ApJ...735..112J} {735, 112}

\bibitem[\protect\citeauthoryear{{Jarrett} et~al.,}{{Jarrett}
  et~al.}{2017}]{Jarrett_2017}
{Jarrett} T.~H.,  et~al., 2017, \mn@doi [\apj] {10.3847/1538-4357/836/2/182},
  \href {http://adsabs.harvard.edu/abs/2017ApJ...836..182J} {836, 182}

\bibitem[\protect\citeauthoryear{{Kodama} \& {Arimoto}}{{Kodama} \&
  {Arimoto}}{1997}]{Kodama_1997}
{Kodama} T.,  {Arimoto} N.,  1997, \aap, \href
  {http://adsabs.harvard.edu/abs/1997A%26A...320...41K} {320, 41}

\bibitem[\protect\citeauthoryear{{Kroupa}}{{Kroupa}}{2001}]{Kroupa_2001}
{Kroupa} P.,  2001, \mn@doi [\mnras] {10.1046/j.1365-8711.2001.04022.x}, \href
  {http://adsabs.harvard.edu/abs/2001MNRAS.322..231K} {322, 231}

\bibitem[\protect\citeauthoryear{{Lang}}{{Lang}}{2014}]{Lang_2014_unWISE}
{Lang} D.,  2014, \mn@doi [\aj] {10.1088/0004-6256/147/5/108}, \href
  {http://cdsads.u-strasbg.fr/abs/2014AJ....147..108L} {147, 108}

\bibitem[\protect\citeauthoryear{{Lang}, {Hogg}  \& {Schlegel}}{{Lang}
  et~al.}{2014}]{Lang_2014_forced_photometry}
{Lang} D.,  {Hogg} D.~W.,   {Schlegel} D.~J.,  2014, preprint, \href
  {http://cdsads.u-strasbg.fr/abs/2014arXiv1410.7397L} {} (\mn@eprint {arXiv}
  {1410.7397})

\bibitem[\protect\citeauthoryear{{Lavoie} et~al.,}{{Lavoie}
  et~al.}{2016}]{Lavoie_2016}
{Lavoie} S.,  et~al., 2016, \mn@doi [\mnras] {10.1093/mnras/stw1906}, \href
  {http://cdsads.u-strasbg.fr/abs/2016MNRAS.462.4141L} {462, 4141}

\bibitem[\protect\citeauthoryear{{Leja}, {Johnson}, {Conroy}  \& {van
  Dokkum}}{{Leja} et~al.}{2018}]{Leja_2018}
{Leja} J.,  {Johnson} B.~D.,  {Conroy} C.,   {van Dokkum} P.,  2018, \mn@doi
  [\apj] {10.3847/1538-4357/aaa8db}, \href
  {http://adsabs.harvard.edu/abs/2018ApJ...854...62L} {854, 62}

\bibitem[\protect\citeauthoryear{{Lidman} et~al.,}{{Lidman}
  et~al.}{2012}]{Lidman_2012}
{Lidman} C.,  et~al., 2012, \mn@doi [\mnras]
  {10.1111/j.1365-2966.2012.21984.x}, \href
  {http://adsabs.harvard.edu/abs/2012MNRAS.427..550L} {427, 550}

\bibitem[\protect\citeauthoryear{{Lidman} et~al.,}{{Lidman}
  et~al.}{2013}]{Lidman_2013}
{Lidman} C.,  et~al., 2013, \mn@doi [\mnras] {10.1093/mnras/stt777}, \href
  {http://adsabs.harvard.edu/abs/2013MNRAS.433..825L} {433, 825}

\bibitem[\protect\citeauthoryear{{Lin}, {Brodwin}, {Gonzalez}, {Bode},
  {Eisenhardt}, {Stanford}  \& {Vikhlinin}}{{Lin} et~al.}{2013}]{Lin_2013}
{Lin} Y.-T.,  {Brodwin} M.,  {Gonzalez} A.~H.,  {Bode} P.,  {Eisenhardt}
  P.~R.~M.,  {Stanford} S.~A.,   {Vikhlinin} A.,  2013, \mn@doi [\apj]
  {10.1088/0004-637X/771/1/61}, \href
  {http://cdsads.u-strasbg.fr/abs/2013ApJ...771...61L} {771, 61}

\bibitem[\protect\citeauthoryear{{Liu}, {Mao}  \& {Meng}}{{Liu}
  et~al.}{2012}]{Liu_2012}
{Liu} F.~S.,  {Mao} S.,   {Meng} X.~M.,  2012, \mn@doi [\mnras]
  {10.1111/j.1365-2966.2012.20886.x}, \href
  {http://cdsads.u-strasbg.fr/abs/2012MNRAS.423..422L} {423, 422}

\bibitem[\protect\citeauthoryear{{Loubser} \&
  {S{\'a}nchez-Bl{\'a}zquez}}{{Loubser} \&
  {S{\'a}nchez-Bl{\'a}zquez}}{2012}]{Loubser_2012}
{Loubser} S.~I.,  {S{\'a}nchez-Bl{\'a}zquez} P.,  2012, \mn@doi [\mnras]
  {10.1111/j.1365-2966.2012.21079.x}, \href
  {http://cdsads.u-strasbg.fr/abs/2012MNRAS.425..841L} {425, 841}

\bibitem[\protect\citeauthoryear{{Loubser}, {S{\'a}nchez-Bl{\'a}zquez},
  {Sansom}  \& {Soechting}}{{Loubser} et~al.}{2009}]{Loubser_2009}
{Loubser} S.~I.,  {S{\'a}nchez-Bl{\'a}zquez} P.,  {Sansom} A.~E.,   {Soechting}
  I.~K.,  2009, \mn@doi [\mnras] {10.1111/j.1365-2966.2009.15171.x}, \href
  {http://cdsads.u-strasbg.fr/abs/2009MNRAS.398..133L} {398, 133}

\bibitem[\protect\citeauthoryear{{Loubser}, {Babul}, {Hoekstra}, {Mahdavi},
  {Donahue}, {Bildfell}  \& {Voit}}{{Loubser} et~al.}{2016}]{Loubser_2016}
{Loubser} S.~I.,  {Babul} A.,  {Hoekstra} H.,  {Mahdavi} A.,  {Donahue} M.,
  {Bildfell} C.,   {Voit} G.~M.,  2016, \mn@doi [\mnras]
  {10.1093/mnras/stv2784}, \href
  {http://cdsads.u-strasbg.fr/abs/2016MNRAS.456.1565L} {456, 1565}

\bibitem[\protect\citeauthoryear{{Mancone} \& {Gonzalez}}{{Mancone} \&
  {Gonzalez}}{2012}]{Mancone_2012}
{Mancone} C.~L.,  {Gonzalez} A.~H.,  2012, \mn@doi [\pasp] {10.1086/666502},
  \href {http://cdsads.u-strasbg.fr/abs/2012PASP..124..606M} {124, 606}

\bibitem[\protect\citeauthoryear{{Maraston}, {Nieves Colmen{\'a}rez}, {Bender}
  \& {Thomas}}{{Maraston} et~al.}{2009}]{Maraston_2009}
{Maraston} C.,  {Nieves Colmen{\'a}rez} L.,  {Bender} R.,   {Thomas} D.,  2009,
  \mn@doi [\aap] {10.1051/0004-6361:20066907}, \href
  {http://adsabs.harvard.edu/abs/2009A%26A...493..425M} {493, 425}

\bibitem[\protect\citeauthoryear{{Maraston} et~al.,}{{Maraston}
  et~al.}{2013}]{Maraston_2013}
{Maraston} C.,  et~al., 2013, \mn@doi [\mnras] {10.1093/mnras/stt1424}, \href
  {http://cdsads.u-strasbg.fr/abs/2013MNRAS.435.2764M} {435, 2764}

\bibitem[\protect\citeauthoryear{{McCarthy}, {Balogh}, {Babul}, {Poole}  \&
  {Horner}}{{McCarthy} et~al.}{2004}]{McCarthy_2004}
{McCarthy} I.~G.,  {Balogh} M.~L.,  {Babul} A.,  {Poole} G.~B.,   {Horner}
  D.~J.,  2004, \mn@doi [\apj] {10.1086/423267}, \href
  {http://cdsads.u-strasbg.fr/abs/2004ApJ...613..811M} {613, 811}

\bibitem[\protect\citeauthoryear{{McCarthy}, {Babul}, {Bower}  \&
  {Balogh}}{{McCarthy} et~al.}{2008}]{McCarthy_2008}
{McCarthy} I.~G.,  {Babul} A.,  {Bower} R.~G.,   {Balogh} M.~L.,  2008, \mn@doi
  [\mnras] {10.1111/j.1365-2966.2008.13141.x}, \href
  {http://cdsads.u-strasbg.fr/abs/2008MNRAS.386.1309M} {386, 1309}

\bibitem[\protect\citeauthoryear{{Muzzin} et~al.,}{{Muzzin}
  et~al.}{2009}]{Muzzin_2009}
{Muzzin} A.,  et~al., 2009, \mn@doi [\apj] {10.1088/0004-637X/698/2/1934},
  \href {http://cdsads.u-strasbg.fr/abs/2009ApJ...698.1934M} {698, 1934}

\bibitem[\protect\citeauthoryear{{Muzzin} et~al.,}{{Muzzin}
  et~al.}{2012}]{Muzzin_2012}
{Muzzin} A.,  et~al., 2012, \mn@doi [\apj] {10.1088/0004-637X/746/2/188}, \href
  {http://adsabs.harvard.edu/abs/2012ApJ...746..188M} {746, 188}

\bibitem[\protect\citeauthoryear{{Oliva-Altamirano} et~al.,}{{Oliva-Altamirano}
  et~al.}{2014}]{Oliva_Altamirano_2014}
{Oliva-Altamirano} P.,  et~al., 2014, \mn@doi [\mnras] {10.1093/mnras/stu277},
  \href {http://cdsads.u-strasbg.fr/abs/2014MNRAS.440..762O} {440, 762}

\bibitem[\protect\citeauthoryear{{Pipino}, {Kaviraj}, {Bildfell}, {Babul},
  {Hoekstra}  \& {Silk}}{{Pipino} et~al.}{2009}]{Pipino_2009}
{Pipino} A.,  {Kaviraj} S.,  {Bildfell} C.,  {Babul} A.,  {Hoekstra} H.,
  {Silk} J.,  2009, \mn@doi [\mnras] {10.1111/j.1365-2966.2009.14534.x}, \href
  {http://cdsads.u-strasbg.fr/abs/2009MNRAS.395..462P} {395, 462}

\bibitem[\protect\citeauthoryear{{Pipino}, {Szabo}, {Pierpaoli}, {MacKenzie}
  \& {Dong}}{{Pipino} et~al.}{2011}]{Pipino_2011}
{Pipino} A.,  {Szabo} T.,  {Pierpaoli} E.,  {MacKenzie} S.~M.,   {Dong} F.,
  2011, \mn@doi [\mnras] {10.1111/j.1365-2966.2011.19444.x}, \href
  {http://cdsads.u-strasbg.fr/abs/2011MNRAS.417.2817P} {417, 2817}

\bibitem[\protect\citeauthoryear{{Popescu}, {Tuffs}, {Dopita}, {Fischera},
  {Kylafis}  \& {Madore}}{{Popescu} et~al.}{2011}]{Popescu_2011}
{Popescu} C.~C.,  {Tuffs} R.~J.,  {Dopita} M.~A.,  {Fischera} J.,  {Kylafis}
  N.~D.,   {Madore} B.~F.,  2011, \mn@doi [\aap] {10.1051/0004-6361/201015217},
  \href {http://adsabs.harvard.edu/abs/2011A%26A...527A.109P} {527, A109}

\bibitem[\protect\citeauthoryear{{Rafferty}, {McNamara}  \&
  {Nulsen}}{{Rafferty} et~al.}{2008}]{Rafferty_2008}
{Rafferty} D.~A.,  {McNamara} B.~R.,   {Nulsen} P.~E.~J.,  2008, \mn@doi [\apj]
  {10.1086/591240}, \href {http://adsabs.harvard.edu/abs/2008ApJ...687..899R}
  {687, 899}

\bibitem[\protect\citeauthoryear{{Rudick}, {Mihos}, {Harding}, {Feldmeier},
  {Janowiecki}  \& {Morrison}}{{Rudick} et~al.}{2010}]{Rudick_2010}
{Rudick} C.~S.,  {Mihos} J.~C.,  {Harding} P.,  {Feldmeier} J.~J.,
  {Janowiecki} S.,   {Morrison} H.~L.,  2010, \mn@doi [\apj]
  {10.1088/0004-637X/720/1/569}, \href
  {http://cdsads.u-strasbg.fr/abs/2010ApJ...720..569R} {720, 569}

\bibitem[\protect\citeauthoryear{{Runge} \& {Yan}}{{Runge} \&
  {Yan}}{2018}]{Runge_2018}
{Runge} J.,  {Yan} H.,  2018, \mn@doi [\apj] {10.3847/1538-4357/aaa020}, \href
  {http://cdsads.u-strasbg.fr/abs/2018ApJ...853...47R} {853, 47}

\bibitem[\protect\citeauthoryear{{Rykoff} et~al.,}{{Rykoff}
  et~al.}{2014}]{Rykoff_2014}
{Rykoff} E.~S.,  et~al., 2014, \mn@doi [\apj] {10.1088/0004-637X/785/2/104},
  \href {http://cdsads.u-strasbg.fr/abs/2014ApJ...785..104R} {785, 104}

\bibitem[\protect\citeauthoryear{{Salpeter}}{{Salpeter}}{1955}]{Salpeter_1955}
{Salpeter} E.~E.,  1955, \mn@doi [\apj] {10.1086/145971}, \href
  {http://cdsads.u-strasbg.fr/abs/1955ApJ...121..161S} {121, 161}

\bibitem[\protect\citeauthoryear{{Schechter}}{{Schechter}}{1976}]{Schechter_1976}
{Schechter} P.,  1976, \mn@doi [\apj] {10.1086/154079}, \href
  {http://cdsads.u-strasbg.fr/abs/1976ApJ...203..297S} {203, 297}

\bibitem[\protect\citeauthoryear{{Schlegel}, {Finkbeiner}  \&
  {Davis}}{{Schlegel} et~al.}{1998}]{Schlegel_1998}
{Schlegel} D.~J.,  {Finkbeiner} D.~P.,   {Davis} M.,  1998, \mn@doi [\apj]
  {10.1086/305772}, \href {http://adsabs.harvard.edu/abs/1998ApJ...500..525S}
  {500, 525}

\bibitem[\protect\citeauthoryear{{Sereno}, {Covone}, {Izzo}, {Ettori}, {Coupon}
   \& {Lieu}}{{Sereno} et~al.}{2017}]{Sereno2017}
{Sereno} M.,  {Covone} G.,  {Izzo} L.,  {Ettori} S.,  {Coupon} J.,   {Lieu} M.,
   2017, \mn@doi [\mnras] {10.1093/mnras/stx2085}, \href
  {http://adsabs.harvard.edu/abs/2017MNRAS.472.1946S} {472, 1946}

\bibitem[\protect\citeauthoryear{{Shankar} et~al.,}{{Shankar}
  et~al.}{2015}]{Shankar_2015}
{Shankar} F.,  et~al., 2015, \mn@doi [\apj] {10.1088/0004-637X/802/2/73}, \href
  {http://cdsads.u-strasbg.fr/abs/2015ApJ...802...73S} {802, 73}

\bibitem[\protect\citeauthoryear{{Speagle}, {Steinhardt}, {Capak}  \&
  {Silverman}}{{Speagle} et~al.}{2014}]{Speagle_2014}
{Speagle} J.~S.,  {Steinhardt} C.~L.,  {Capak} P.~L.,   {Silverman} J.~D.,
  2014, \mn@doi [\apjs] {10.1088/0067-0049/214/2/15}, \href
  {http://adsabs.harvard.edu/abs/2014ApJS..214...15S} {214, 15}

\bibitem[\protect\citeauthoryear{{Stern} et~al.,}{{Stern}
  et~al.}{2012}]{Stern_2012}
{Stern} D.,  et~al., 2012, \mn@doi [\apj] {10.1088/0004-637X/753/1/30}, \href
  {http://cdsads.u-strasbg.fr/abs/2012ApJ...753...30S} {753, 30}

\bibitem[\protect\citeauthoryear{{Stott}, {Edge}, {Smith}, {Swinbank}  \&
  {Ebeling}}{{Stott} et~al.}{2008}]{Stott_2008}
{Stott} J.~P.,  {Edge} A.~C.,  {Smith} G.~P.,  {Swinbank} A.~M.,   {Ebeling}
  H.,  2008, \mn@doi [\mnras] {10.1111/j.1365-2966.2007.12807.x}, \href
  {http://cdsads.u-strasbg.fr/abs/2008MNRAS.384.1502S} {384, 1502}

\bibitem[\protect\citeauthoryear{{Strateva} et~al.,}{{Strateva}
  et~al.}{2001}]{Strateva_2001}
{Strateva} I.,  et~al., 2001, \mn@doi [\aj] {10.1086/323301}, \href
  {http://cdsads.u-strasbg.fr/abs/2001AJ....122.1861S} {122, 1861}

\bibitem[\protect\citeauthoryear{{Szabo}, {Pierpaoli}, {Dong}, {Pipino}  \&
  {Gunn}}{{Szabo} et~al.}{2011}]{Szabo_2011}
{Szabo} T.,  {Pierpaoli} E.,  {Dong} F.,  {Pipino} A.,   {Gunn} J.,  2011,
  \mn@doi [\apj] {10.1088/0004-637X/736/1/21}, \href
  {http://cdsads.u-strasbg.fr/abs/2011ApJ...736...21S} {736, 21}

\bibitem[\protect\citeauthoryear{{Thomas}, {Maraston}, {Bender}  \& {Mendes de
  Oliveira}}{{Thomas} et~al.}{2005}]{Thomas_2005}
{Thomas} D.,  {Maraston} C.,  {Bender} R.,   {Mendes de Oliveira} C.,  2005,
  \mn@doi [\apj] {10.1086/426932}, \href
  {http://cdsads.u-strasbg.fr/abs/2005ApJ...621..673T} {621, 673}

\bibitem[\protect\citeauthoryear{{Tonini}, {Bernyk}, {Croton}, {Maraston}  \&
  {Thomas}}{{Tonini} et~al.}{2012}]{Tonini_2012}
{Tonini} C.,  {Bernyk} M.,  {Croton} D.,  {Maraston} C.,   {Thomas} D.,  2012,
  \mn@doi [\apj] {10.1088/0004-637X/759/1/43}, \href
  {http://cdsads.u-strasbg.fr/abs/2012ApJ...759...43T} {759, 43}

\bibitem[\protect\citeauthoryear{{Tonry}}{{Tonry}}{1987}]{Tonry_1987}
{Tonry} J.~L.,  1987, in {de Zeeuw} P.~T.,  ed.,  IAU Symposium Vol. 127,
  Structure and Dynamics of Elliptical Galaxies. pp 89--96

\bibitem[\protect\citeauthoryear{{Valentinuzzi} et~al.,}{{Valentinuzzi}
  et~al.}{2011}]{Valentinuzzi_2011}
{Valentinuzzi} T.,  et~al., 2011, \mn@doi [\aap] {10.1051/0004-6361/201117522},
  \href {http://adsabs.harvard.edu/abs/2011A%26A...536A..34V} {536, A34}

\bibitem[\protect\citeauthoryear{{Webb} et~al.,}{{Webb}
  et~al.}{2015}]{Webb_2015}
{Webb} T.~M.~A.,  et~al., 2015, \mn@doi [\apj] {10.1088/0004-637X/814/2/96},
  \href {http://cdsads.u-strasbg.fr/abs/2015ApJ...814...96W} {814, 96}

\bibitem[\protect\citeauthoryear{{Webb} et~al.,}{{Webb}
  et~al.}{2017}]{Webb_2017}
{Webb} T.~M.~A.,  et~al., 2017, \mn@doi [\apjl] {10.3847/2041-8213/aa7749},
  \href {http://cdsads.u-strasbg.fr/abs/2017ApJ...844L..17W} {844, L17}

\bibitem[\protect\citeauthoryear{{Wen} \& {Han}}{{Wen} \&
  {Han}}{2015a}]{Wen_2015b}
{Wen} Z.~L.,  {Han} J.~L.,  2015a, \mn@doi [\mnras] {10.1093/mnras/stu2722},
  \href {http://cdsads.u-strasbg.fr/abs/2015MNRAS.448....2W} {448, 2}

\bibitem[\protect\citeauthoryear{{Wen} \& {Han}}{{Wen} \&
  {Han}}{2015b}]{Wen_2015a}
{Wen} Z.~L.,  {Han} J.~L.,  2015b, \mn@doi [\apj]
  {10.1088/0004-637X/807/2/178}, \href
  {http://cdsads.u-strasbg.fr/abs/2015ApJ...807..178W} {807, 178}

\bibitem[\protect\citeauthoryear{{Wen}, {Han}  \& {Liu}}{{Wen}
  et~al.}{2009}]{Wen_2009}
{Wen} Z.~L.,  {Han} J.~L.,   {Liu} F.~S.,  2009, \mn@doi [\apjs]
  {10.1088/0067-0049/183/2/197}, \href
  {http://cdsads.u-strasbg.fr/abs/2009ApJS..183..197W} {183, 197}

\bibitem[\protect\citeauthoryear{{Wen}, {Han}  \& {Liu}}{{Wen}
  et~al.}{2012}]{Wen_2012}
{Wen} Z.~L.,  {Han} J.~L.,   {Liu} F.~S.,  2012, \mn@doi [\apjs]
  {10.1088/0067-0049/199/2/34}, \href
  {http://cdsads.u-strasbg.fr/abs/2012ApJS..199...34W} {199, 34}

\bibitem[\protect\citeauthoryear{{Whiley} et~al.,}{{Whiley}
  et~al.}{2008}]{Whiley_2008}
{Whiley} I.~M.,  et~al., 2008, \mn@doi [\mnras]
  {10.1111/j.1365-2966.2008.13324.x}, \href
  {http://cdsads.u-strasbg.fr/abs/2008MNRAS.387.1253W} {387, 1253}

\bibitem[\protect\citeauthoryear{{White}}{{White}}{1976}]{White_1976}
{White} S.~D.~M.,  1976, \mn@doi [\mnras] {10.1093/mnras/174.1.19}, \href
  {http://cdsads.u-strasbg.fr/abs/1976MNRAS.174...19W} {174, 19}

\bibitem[\protect\citeauthoryear{{Wilson} et~al.,}{{Wilson}
  et~al.}{2009}]{Wilson_2009}
{Wilson} G.,  et~al., 2009, \mn@doi [\apj] {10.1088/0004-637X/698/2/1943},
  \href {http://cdsads.u-strasbg.fr/abs/2009ApJ...698.1943W} {698, 1943}

\bibitem[\protect\citeauthoryear{{Worthey}}{{Worthey}}{1994}]{Worthey_1994}
{Worthey} G.,  1994, \mn@doi [\apjs] {10.1086/192096}, \href
  {http://adsabs.harvard.edu/abs/1994ApJS...95..107W} {95, 107}

\bibitem[\protect\citeauthoryear{{Wright} et~al.,}{{Wright}
  et~al.}{2010}]{Wright_2010}
{Wright} E.~L.,  et~al., 2010, \mn@doi [\aj] {10.1088/0004-6256/140/6/1868},
  \href {http://cdsads.u-strasbg.fr/abs/2010AJ....140.1868W} {140, 1868}

\bibitem[\protect\citeauthoryear{{York} et~al.,}{{York}
  et~al.}{2000}]{York_2000}
{York} D.~G.,  et~al., 2000, \mn@doi [\aj] {10.1086/301513}, \href
  {http://cdsads.u-strasbg.fr/abs/2000AJ....120.1579Y} {120, 1579}

\bibitem[\protect\citeauthoryear{{Yuan}, {Han}  \& {Wen}}{{Yuan}
  et~al.}{2016}]{Yuan_2016}
{Yuan} Z.~S.,  {Han} J.~L.,   {Wen} Z.~L.,  2016, \mn@doi [\mnras]
  {10.1093/mnras/stw1125}, \href
  {http://cdsads.u-strasbg.fr/abs/2016MNRAS.460.3669Y} {460, 3669}

\bibitem[\protect\citeauthoryear{{Zhao}, {Arag{\'o}n-Salamanca}  \&
  {Conselice}}{{Zhao} et~al.}{2015}]{Zhao_2015b}
{Zhao} D.,  {Arag{\'o}n-Salamanca} A.,   {Conselice} C.~J.,  2015, \mn@doi
  [\mnras] {10.1093/mnras/stv1940}, \href
  {http://cdsads.u-strasbg.fr/abs/2015MNRAS.453.4444Z} {453, 4444}

\bibitem[\protect\citeauthoryear{{de Filippis}, {Paolillo}, {Longo}, {La
  Barbera}, {de Carvalho}  \& {Gal}}{{de Filippis}
  et~al.}{2011}]{De_Filippis_2011}
{de Filippis} E.,  {Paolillo} M.,  {Longo} G.,  {La Barbera} F.,  {de Carvalho}
  R.~R.,   {Gal} R.,  2011, \mn@doi [\mnras]
  {10.1111/j.1365-2966.2011.18596.x}, \href
  {http://cdsads.u-strasbg.fr/abs/2011MNRAS.414.2771D} {414, 2771}

\bibitem[\protect\citeauthoryear{{de Vaucouleurs}}{{de
  Vaucouleurs}}{1948}]{DeVauc_1948}
{de Vaucouleurs} G.,  1948, Annales d'Astrophysique, \href
  {http://adsabs.harvard.edu/abs/1948AnAp...11..247D} {11, 247}

\bibitem[\protect\citeauthoryear{{van Dokkum}, {Franx}, {Kelson},
  {Illingworth}, {Fisher}  \& {Fabricant}}{{van Dokkum}
  et~al.}{1998}]{van_Dokkum_1998}
{van Dokkum} P.~G.,  {Franx} M.,  {Kelson} D.~D.,  {Illingworth} G.~D.,
  {Fisher} D.,   {Fabricant} D.,  1998, \mn@doi [\apj] {10.1086/305762}, \href
  {http://adsabs.harvard.edu/abs/1998ApJ...500..714V} {500, 714}

\bibitem[\protect\citeauthoryear{{von der Linden}, {Best}, {Kauffmann}  \&
  {White}}{{von der Linden} et~al.}{2007}]{von_der_Linden_2007}
{von der Linden} A.,  {Best} P.~N.,  {Kauffmann} G.,   {White} S.~D.~M.,  2007,
  \mn@doi [\mnras] {10.1111/j.1365-2966.2007.11940.x}, \href
  {http://cdsads.u-strasbg.fr/abs/2007MNRAS.379..867V} {379, 867}

\makeatother
\end{thebibliography}




\appendix

\section{WISE-detecetd BCGs}

\begin{figure*}
  \centering
	\includegraphics[width=0.9\textwidth, trim=0.0cm 0.0cm 0.0cm 0.0cm, clip, page=1]{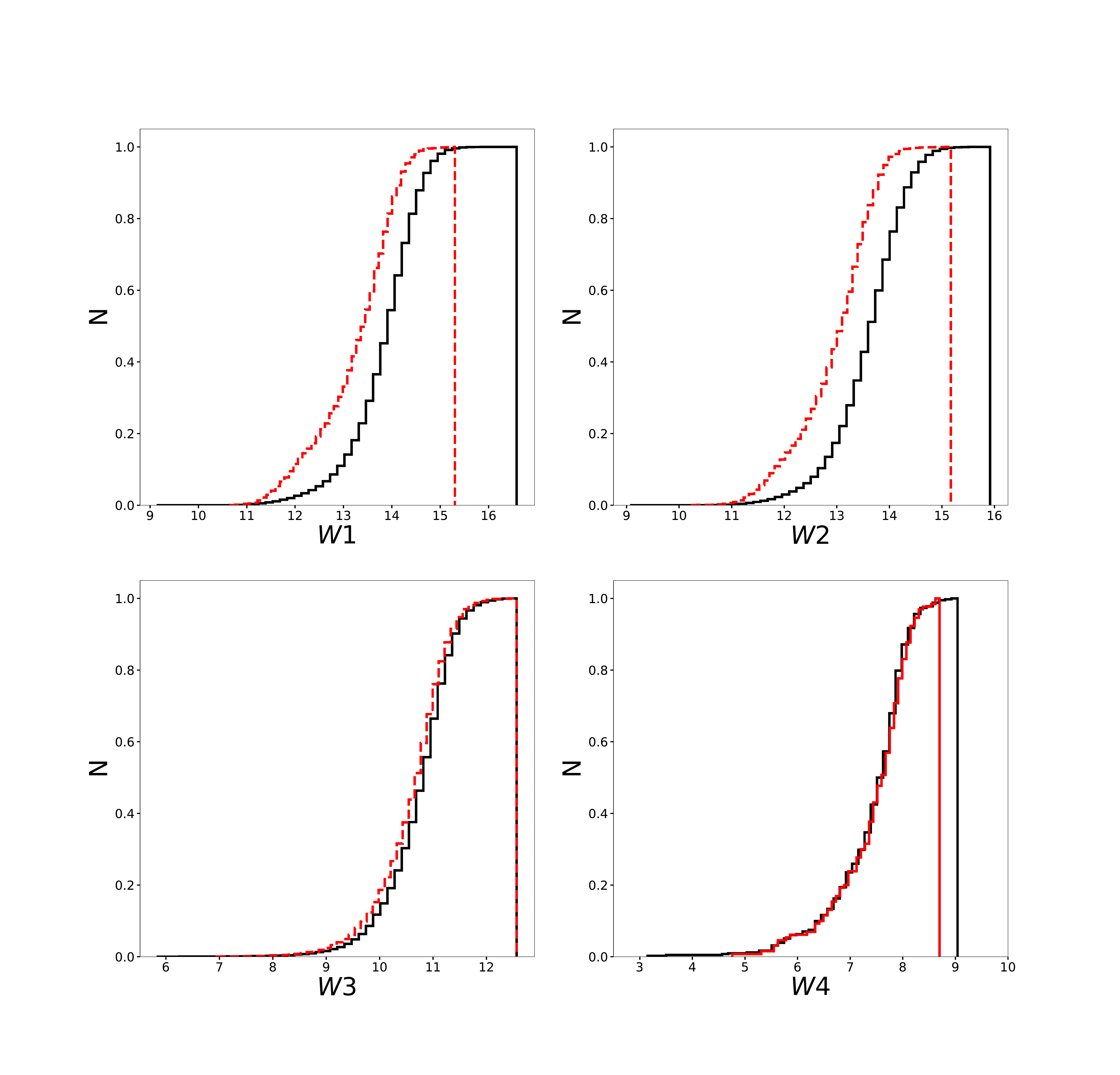}
	\caption{The cumulative $W1$, $W2$, $W3$ and $W4$ magnitude distributions of the BCGs detected in WISE at $0.05 \leq z < 0.35$. The black solid histograms represent the detections in each WISE band, while the red dashed histograms are are the magnitude cumulative distributions for the galaxies simultaneously detected in $W1$, $W2$ and $W3$ for the top-left, top-righte and bottom-left panels and in $W1$, $W2$, $W3$ and $W4$ for the bottom-right panel.}
\label{fig:plot_WISE_histograms}
\end{figure*}

The present Appendix presents the magnitude distributions in the four WISE bands for the sample of BCGs studied in this paper. Each plot presents the cumulative distributions of the $W1$, $W2$, $W3$, and $W4$ magnitudes for the objects detected in the AllWISE catalogue. Black histograms represent the distributions of single band detections at $0.05 \leq z_{phot}$ < 0.35, while red histograms are the magnitude distributions in the same bands after the cuts in cluster halo mass, optical photometry quality and photometric redshift quality described in Section \ref{optical_data} are applied. In the cases of $W1$, $W2$ and $W3$ the red histograms correspond to objects simultaneously detected in these 3 bands, while in the case of $W4$ the red histogram corresponds to detections in all the four bands. Although we did not use $W4$ in this paper we include the magnitude distribtions in this band for completeness. All the histograms are normalised to sum up to 1.

As expected, the sample of objects simultaeously detected in $W1$, $W2$ and $W3$ does not include the faintest $W1$ and $W2$ emitters, and the bottom-left panel shows that this sample is essentially $W3$-selected.


\bsp	
\label{lastpage}
\end{document}